\documentclass[preprint2]{aastex}

\slugcomment{Not to appear in Nonlearned J., 45.}

\shorttitle{Revisiting the fossil group candidates UGC 842 and NGC 6034}
\shortauthors{Lopes de Oliveira et al.}

\begin{document}

\shorttitle{Revisiting the fossil group candidates UGC 842 and NGC 6034}
\shortauthors{Lopes de Oliveira et al.}

\title{Revisiting the fossil group candidates UGC 842 and NGC 6034}

\author{
R. Lopes de Oliveira\altaffilmark{1}, 
E. R. Carrasco\altaffilmark{2},
C. Mendes de Oliveira\altaffilmark{1}, 
D. R. Bortoletto\altaffilmark{1,3},
E. Cypriano\altaffilmark{1},
L. Sodr\'e Jr.\altaffilmark{1},
G. B. Lima Neto\altaffilmark{1}
}

\altaffiltext{1}{Departamento de Astronomia, Instituto de Astronomia, Geof\'isica e Ci\^encias Atmosf\'ericas da Universidade de S\~ao Paulo, Rua do Mat\~ao 1226, Cidade Universit\'aria, 05508-090, Brazil; rlopes@astro.iag.usp.br.}
\altaffiltext{2}{Gemini Observatory, Southern Operations Center, AURA, Casilla 603, La Serena, Chile}
\altaffiltext{3}{Laborat\'orio Nacional de Astrof\'isica, Itajub\'a, Brazil}

\begin{abstract}

We present a new insight on NGC 6034 and UGC 842, two groups of galaxies
previously reported in the literature as being fossil groups. The study is based
on optical photometry and spectroscopy  obtained with the CTIO Blanco telescope
and Sloan Digital Sky Survey archival data.  We
find that NGC 6034 is embedded in a large structure, dominated by three  rich
clusters and other small groups. Its first and next four ranked galaxies have magnitude differences in the $r$ band and projected distances which violate the optical
criteria to classify it as a fossil group.  We confirm
that the UGC 842 group is a fossil group, but with about half the velocity dispersion that is
reported in previous works. The velocity distribution of its galaxies reveals the existence of two structures in its line of sight, one with $\sigma_v$ $\sim$ 223 km\,s$^{-1}$ and another with $\sigma_v$ $\sim$ 235 km\,s$^{-1}$, with a difference in velocity of  $\sim$ 820 km\,s$^{-1}$. The main
structure is dominated by passive galaxies, while these represent $\sim$ 60\% of the second structure. The X-ray temperature for the intragroup medium of a group
with such a velocity dispersion is expected to be $kT$ $\sim$ 0.5--1 keV, against
the observed value of $kT$ $\sim$ 1.9  keV reported in the literature. This
result makes UGC 842 a special case among fossil groups  because (1) it
represents more likely the interaction between two small groups, which warms the 
intragroup medium and/or (2) it could constitute evidence that member galaxies
lost energy in  the process of spiraling toward the group center, and
decreased the velocity dispersion of the  system. As far as we know, UGC 842 is
the first low-mass fossil group studied in detail.

\end{abstract}

\keywords{cosmology: observations, galaxies: clusters: individual: UGC 842, NGC 6034, galaxies: elliptical and lenticular, cD, galaxies: evolution, galaxies: kinematics and dynamics, galaxies: intergalactic medium}

\section{Introduction}

Fossil groups are galaxy systems optically dominated
by an elliptical galaxy immersed in an extended and luminous X-ray
halo (L$_{X, \rm bol}$ $>$ 10$^{42}$ $h^{-2}_{50}$ erg\,s$^{-1}$), in
which the magnitude gap between the two brightest galaxies within
half of the virial radius is greater than 2 in the $r$ band
\citep{Ponman94}.  The scarcity of L$^*$ galaxies and the evidence
for the existence of a ``massive" structure traced by the hot intragroup
gas support the hypothesis of their central galaxies being formed
by merging of luminous galaxies, most likely by dynamical friction
\citep{Cypriano06,MdO06,MdO09}.  

The estimated time scales
for dynamical friction and results of numerical simulations are
consistent with an old age for fossil groups. For example, $N$-body/hydrodynamical
simulations carried out by \citet{DOnghia05} suggest a correlation
between the magnitude gap between the two brightest galaxies and
the formation time of the group. More recently several studies of 
simulated fossil groups were made using the Millennium simulations
\citep{Dariush07,Sales07,Diaz08}  which
corroborated the idea that fossil groups/clusters assembled most of their
masses much earlier than non-fossil systems. 
These studies also predicted a fraction of 3\%--13\%  
of groups with masses greater than 10$^{13}$ M$_{\odot}$ being fossil groups,
depending on the exact range of masses considered.

The true nature of fossil groups has promoted a lively debate in
the past decade.  The absence of a conclusive explanation for the
nature of these objects is mainly due to (1) the lack of a proper
sample for statistical studies, (2) the lack of X-ray data with
sufficient resolution and signal-to-noise (S/N) for a proper study of the group properties,
and (3) the lack of optical spectroscopy of the group members, for
membership confirmation and a study of the kinematic properties of the
known groups.
A growing
number of fossil group candidates has been claimed in the recent
literature. For example, \citet{Santos07} identified 34 potential
candidate fossil groups in a systematic search carried out using
the {\it Sloan Digital Sky Survey} (SDSS). 
More recently, \citet{LaBarbera09} identified
25 other fossil groups.
However, the nature of these newly identified groups still need to be confirmed.
We investigate here two groups pointed in the literature as being
fossil groups: UGC 842 and NGC 6034 \citep{Voevodkin08,Yoshioka04}.

UGC 842 is a bright elliptical galaxy with a heliocentric radial velocity of 13,556$\pm$32 km\,s$^{-1}$ \citep{Huchra99}.
It is a weak radio source detected by the Very
Large Array (VLA) telescope with flux level of S(6 cm) $\sim$ 1.1
mJy \citep{Gioia83}. The UGC 842 group is immersed in an extended halo with a radius of
$\sim$ 4' in the sky, which corresponds to $\sim$ 300 kpc at its
redshift. 
\citet{Gastaldello07} reported an X-ray observation of UGC 842 system
carried out by the {\it Chandra} satellite. They derived a virial radius
and mass of 1272$\pm$220 $h^{-1}_{70}$\,kpc and 12.8$\times$10$^{13}$ $h^{-1}_{70}$\,$M_{\odot}$,
and $r_{500}$ and $M_{500}$ of 634$\pm$85 $h^{-1}_{70}$\,kpc and
7.54$\pm$3.41$\times$10$^{13}$ $h^{-1}_{70}$\,$M_{\odot}$, respectively.  
About 1 year later,
\citet{Voevodkin08} reported a combined optical and X-ray analysis
of this group from SDSS and {\it XMM-Newton} data. According to these
authors, galaxies in the UGC 842 group have $\sigma_v$ $\sim$ 439 km\,s$^{-1}$ 
(from 16 galaxies inside $r$ = 509 $h^{-1}_{71}$\,kpc),
its intragroup gas has a temperature $kT$ of 1.90$\pm$0.30\,keV and
metallicity of 0.34$\pm$0.12\,Z$_{\odot}$, and displays a bolometric
X-ray luminosity of $\sim$ 1.63$\times$10$^{43}$\,$h^{-2}_{71}$\,erg\,s$^{-1}$.
From the {\it XMM-Newton} data, \citet{Voevodkin08} also inferred
$r_{500}$ $\sim$ 509 $h^{-1}_{71}$\,kpc, M$_{\rm gas,500}$ = (3.5$\pm$1.1)$\times$10$^{12}$
$h^{-1}_{71}$\,M$_{\odot}$, and M$_{\rm total,500}$ = (4.03$\pm$0.69)$\times$10$^{13}$
$h^{-1}_{71}$\,M$_{\odot}$. 
The UGC 842 group is classified as a fossil group by these authors.

\begin{deluxetable}{cccccccrrrl}
\tabletypesize{\scriptsize}
\tablecaption{Radial velocities and $r$-mag of the central galaxies of UGC 842 and NGC 6034 \label{tbl:sample}}
\tablenum{1}
\tablecolumns{1}
\tablewidth{0pc}
\tablehead{
\colhead{Group} &
\colhead{Central Galaxy} &
\colhead{Radial Velocity} &
\colhead{1st-ranked $r$-mag} &
\colhead{1st-ranked $R$-mag}\\
& &
(CTIO-Hydra) &
(SDSS) &
(SDSS)
}
\startdata
UGC 842   & J011853.62--010007.2  & 13428$\pm$40 km\,s$^{-1}$ & 13.47 & -23.19\\
NGC 6034  & J160332.08+171155.2   & 10162$\pm$48 km\,s$^{-1}$ & 13.54 & -22.49\\
\enddata
\end{deluxetable}

NGC 6034 is a bright E/S0 radio galaxy member of Abell 2151 and
immersed in the Hercules supercluster \citep{Corwin71,Val78}, with
a heliocentric radial velocity of $\sim$ 10,112 km\,s$^{-1}$ 
\citep{Tarenghi79}. VLA observations at $\lambda$21-cm (H\,{\small I})
reported by \citet{Dickey97} show that NGC 6034 has a head-tail
morphology in the continuum, with two jets, and evidence of gas
falling toward the nucleus with a velocity of about 70 km\,s$^{-1}$
as derived from the detection of shifted narrow absorption line at
10,226$\pm$15 km\,s$^{-1}$. An X-ray observation of the NGC 6034 group with
the {\it Einstein} Observatory shows that $L_{\rm X,0.5-4.5
keV}$ $<$ 1.3$\times$10$^{42}$ $h_{50}^{-2}$ erg\,s$^{-1}$
\citep{Canizares87}.
ASCA observation (1999 August 24) reveals that this group has a relatively cold intragroup medium 
with $kT$ = 0.67$\pm$0.09 keV and abundance $Z$ = 0.08$\pm$0.05
$Z_{\odot}$, affected by a photoelectric absorption equivalent to
$N_{H}$ $\sim$ 3.4$\times$10$^{20}$ cm$^{-2}$, and displaying a bolometric
X-ray luminosity of about 2.8$\times$10$^{43}$ $h_{50}^{-2}$
erg\,s$^{-1}$ \citep{Fukazawa04}.
Curiously, also from ASCA observation,
\citet{Yoshioka04} derived $kT$ = 1.29(+0.48/-0.36) keV, $Z$ =
0.11(+0.39/-0.10) $Z_{\odot}$, $N_{H}$ $\sim$ 3.4$\times$10$^{20}$
cm$^{-2}$, and $L_{\rm X, 0.1-2.4 keV}$ $\sim$ 7.5$\times$10$^{41}$ $h_{75}^{-2}$
erg\,s$^{-1}$.
The main properties of the UGC 842 and NGC 6034 groups are described in Table \ref{tbl:sample}.

We report on new optical photometry and spectroscopy of the groups UGC 842 and
NGC 6034 (hereafter UGC 842 and NGC 6034, for simplicity) and their ``neighborhoods", 
carried out at the CTIO-Blanco 4.0 m telescope, 
which are combined with data from the SDSS Data Release 6 \citep[DR6;][]{Adelman08}. These data are used here to perform an analysis
of the structure and kinematics of these two systems and also allowed an investigation about the nature of the systems as fossil groups (confirmed in the case of UGC 842 and
not confirmed for NGC 6034).
When needed, we adopt a lambda cold dark matter ($\Lambda$CDM) cosmology with $H_{0}$ = 70 km\,s$^{-1}$\,Mpc$^{-1}$, $\Omega_{M}$ = 0.3, and $\Omega_{\Lambda}$ = 0.7.

\section{Observations: Photometry and Spectroscopy}

Images in the $B$ and $R$ bands of UGC 842 and NGC
6034 were obtained at the CTIO-Blanco 4.0 m telescope on 2005 August 25.  
The images were taken with
the mosaic camera, covering a region (after trimming of the edges) of approximately
$\sim$ 38$\times$38 arcmin$^2$ (equivalent to about
2$\times$2 Mpc$^2$ and 1.5$\times$1.5 Mpc$^2$ at the redshift of each of
these groups, respectively).

A total of five mosaic images were obtained for UGC 842, in the $B$ band,
with exposure times of 600\,s each, and other five images of 360\,s
each in the $R$ band, with a seeing of about 1 arcsec in the $R$ band
and 1.1 arcsec in the $B$ band.  For NGC 6034, only one image was
taken for each filter, with exposures times of 200\,s and seeing
of 1.6 arcsec in the $R$ band and 2.1 arcsec in the $B$ band.  

The galaxies in the area around NGC 6034 and UGC 842 were observed spectroscopically 
on 2006 August 13 UT with the Hydra-CTIO multi-object spectrograph \citep{Barden98} at the CTIO Blanco 4 m telescope in Chile.
NGC 6034 was also observed on 2007 April 14, with the same equipment.
The observations were performed
during dark/gray times with a good atmospheric transparency in
2007 April, and under bright sky with some cirrus in 2006 August, 
in both epochs with a seeing that on average varied between 0.9'' 
and 1.2'' (DIMM monitor). The observation of UGC 842 had in general poor S/N due to the 
proximity of the Moon in the field -- we were able 
to extract only 20 out of a total of 84 spectra observed.

The spectra were acquired using two different setups. For the data
taken on 2006 August, we used the KPGL3 grating over the wavelength
range 3960--6960\,\AA, centered on 5278\,\AA, which provided a
spectral resolution of $\sim$ 4\,\AA, and a dispersion of 1.39\,\AA\,pixel$^{-1}$.
In 2007 April, we used the KPGL2 grating over the wavelength range
3450--8242\,\AA, centered on 5845\,\AA, which provided a spectral
resolution of $\sim$ 6.5\,\AA, and dispersion of 2.33\,\AA\,pixel$^{-1}$. To
avoid second order contamination above 8000\,\AA, the blocking filter
GG385 was used. All spectra were imaged with the 400 nm Bench Schmidt
camera onto a SITe 2k$\times$4k CCD, with a binning of 2 pixels
in the spectral direction. 
Total exposure times
of 2 hr (4$\times$30 minutes) and 1.5 hr (3$\times$30 minutes)
were used for the objects observed in 2006 August and 2007 April,
respectively.

\subsection {Data Reduction and Analysis}

\subsubsection{Photometry}

The images were processed using standard reduction techniques
within IRAF\footnote{
IRAF is distributed by the National Optical Astronomy Observatories,
    which are operated by the Association of Universities for Research
    in Astronomy, Inc., under cooperative agreement with the National
    Science Foundation.}.
After correcting them by bias and flat field, we derived positions and
magnitudes for all identified objects using Sextractor. Flux calibration
was performed using magnitudes of common stars in the field with the SDSS
database, given that
our nights were not photometric. The probable galaxies were
identified based on the {\it class\_star} parameter (less than 0.7). From the
number counts of galaxies in the whole field we estimate that the photometry
is complete down to about 21.5 mag in the $R$ band for UGC 842 and NGC 6034.

\subsubsection{Spectroscopy}

The spectroscopic data were reduced using standard procedures in IRAF. In summary,
all science exposures, comparison lamps (He/Ne/Ar), spectroscopic flats (taken before or after each object spectrum) and
the so-called ``milk flats''\footnote{Milk flats are calibrations obtained once per run, during the afternoon, by opening the dome and placing a dispersing filter in front of the row of fibers, in the spectrograph.} were bias/overscan subtracted and trimmed using the
IRAF task CCDPROC. The ``milk flats" were then combined and spectral
shapes in $x$- and $y$-direction were removed using the IRAF task FIT1D.
The resultant image was then filtered by using a median filter and
normalized to one. 
The science exposures and spectroscopic flats were then divided
by the processed ``milk flats" in order to reduce
spectral noise in the images.

The spectra were extracted with the IRAF DOHYDRA task. 
Dome flats were used to flat field the
individual fibers, while twilight flats were used for fiber-to-fiber
throughput correction. 
The spectra were then wavelength calibrated.
The residual values in the wavelength solution
for 20--30 points using a 4th or 5th-order Chebyshev polynomial
typically yielded $rms$ values of $\sim$ 0.20--0.30 \AA. Finally,
the average sky spectrum was subtracted from each object spectrum 
using typically 12 sky fiber spectra per field. The
final one-dimensional object spectra were then combined by their
average value.

\subsubsection{Radial velocities}
\label{sct:rv}

In order to measure radial velocities we first inspected the spectra to
search for obvious absorption and emission features characteristic
of early- and late-type galaxy populations. 
For galaxies with clear
emission lines, the IRAF task RVIDLINE was used employing a
line-by-line Gaussian fit to measure the radial velocity 
(this was done for only one galaxy belonging to UGC 842 and 11 galaxies in NGC 6034). 
The
residuals of the average velocity shifts of all measurements were
used to estimate the errors. 
The velocities for absorption-line and
for emission-line systems with clear absorption lines (6 out of 11 galaxies belonging to NGC 6034), were calculated using 
the cross-correlation technique \citep{Tonry79}. The spectra were 
cross-correlated with high S/N templates using the task FXCOR inside IRAF.
The
detected narrow cross-correlation peaks were fitted by a Gaussian,
with errors given by the $R$-statistic in which $\sigma_v$ = (3/8)($w$/(1
+ $R$)) \citep{Tonry79}, where $w$ is the FWHM of the correlation
peak and $R$ is the ratio of the correlation peak height to the
amplitude of the antisymmetric noise.
This $R$ value was used as
a reliability factor of the quality of the measured velocity. For
$R$ $>$ 3.5, the resulting velocity was that associated to the
template which produced the lowest error. For galaxies with $R$ $<$
3.5, we looked for absorption features like Ca\,{\small II} and $G$ band in the
spectra, and performed a line-by-line Gaussian fit using the package
RVIDLINE. The resulting values were then compared with the velocities
given by cross-correlation. In all cases the agreement between the
two procedures was good.
The measured velocities are shown in Tables \ref{tbl:ugcvel} and \ref{tbl:ngcvel} 
(see Section \ref{sct:veldist}) for galaxies in the field of UGC 842 and NGC 6034, respectively.

\subsection{The SDSS catalog}
\label{sct:sdsscat}

We used the SDSS DR6 to search for galaxies in the region around NGC 6034 and UGC 842 with
three objectives: 
(1) to find missing galaxies that were not observed
spectroscopically in our observations in the CTIO; 
(2) to study the environment in a large area around the groups 
and to minimize the contamination produced by other nearby structures
in our analysis (in Section \ref{sct:envngc}); and
(3) to find any systematic differences that may exist between SDSS and our data.

\begin{figure*}[t!]
\centerline{
\includegraphics[angle=-90,width=65mm]{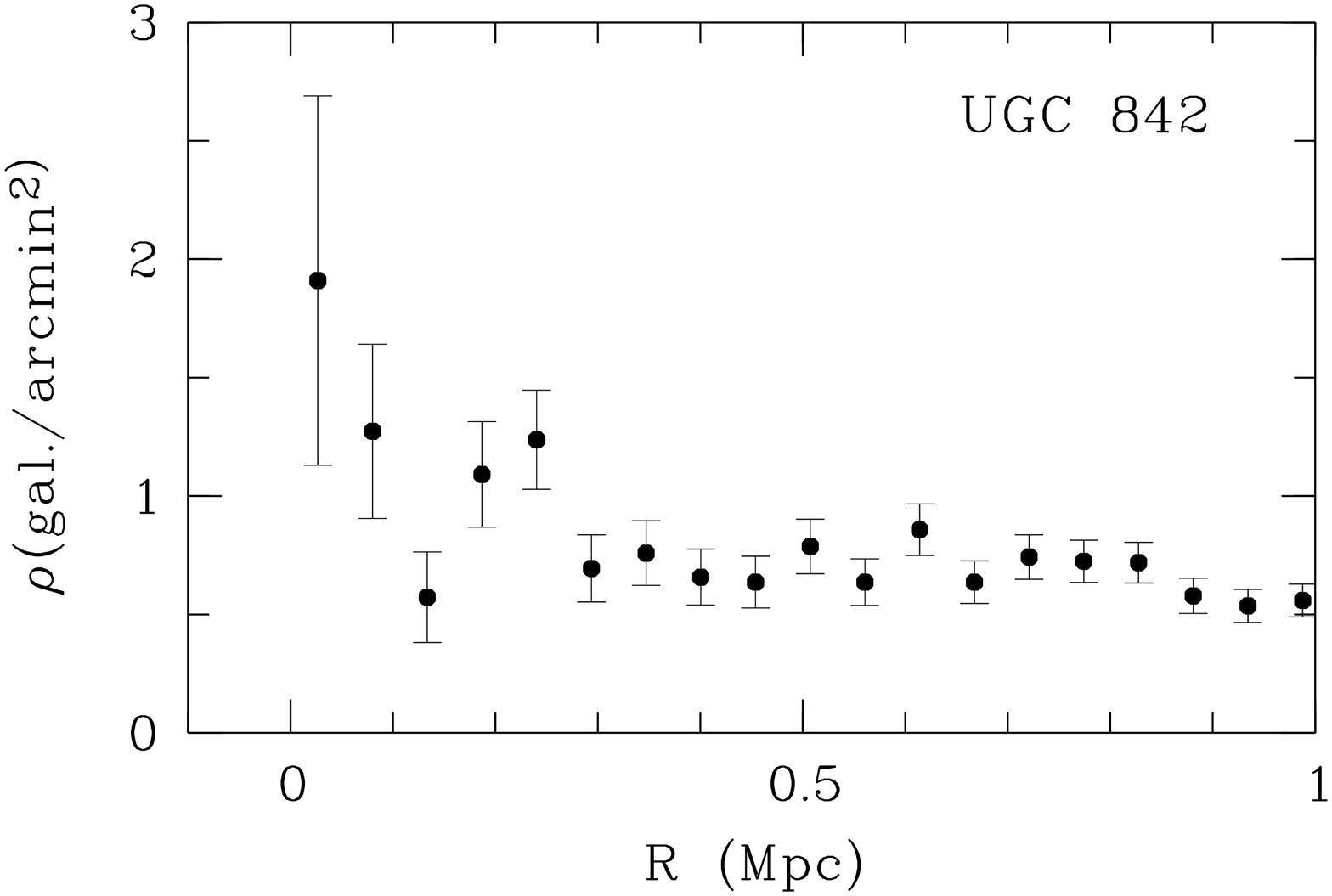}~~~~~~~~ 
\includegraphics[angle=-90,width=65mm]{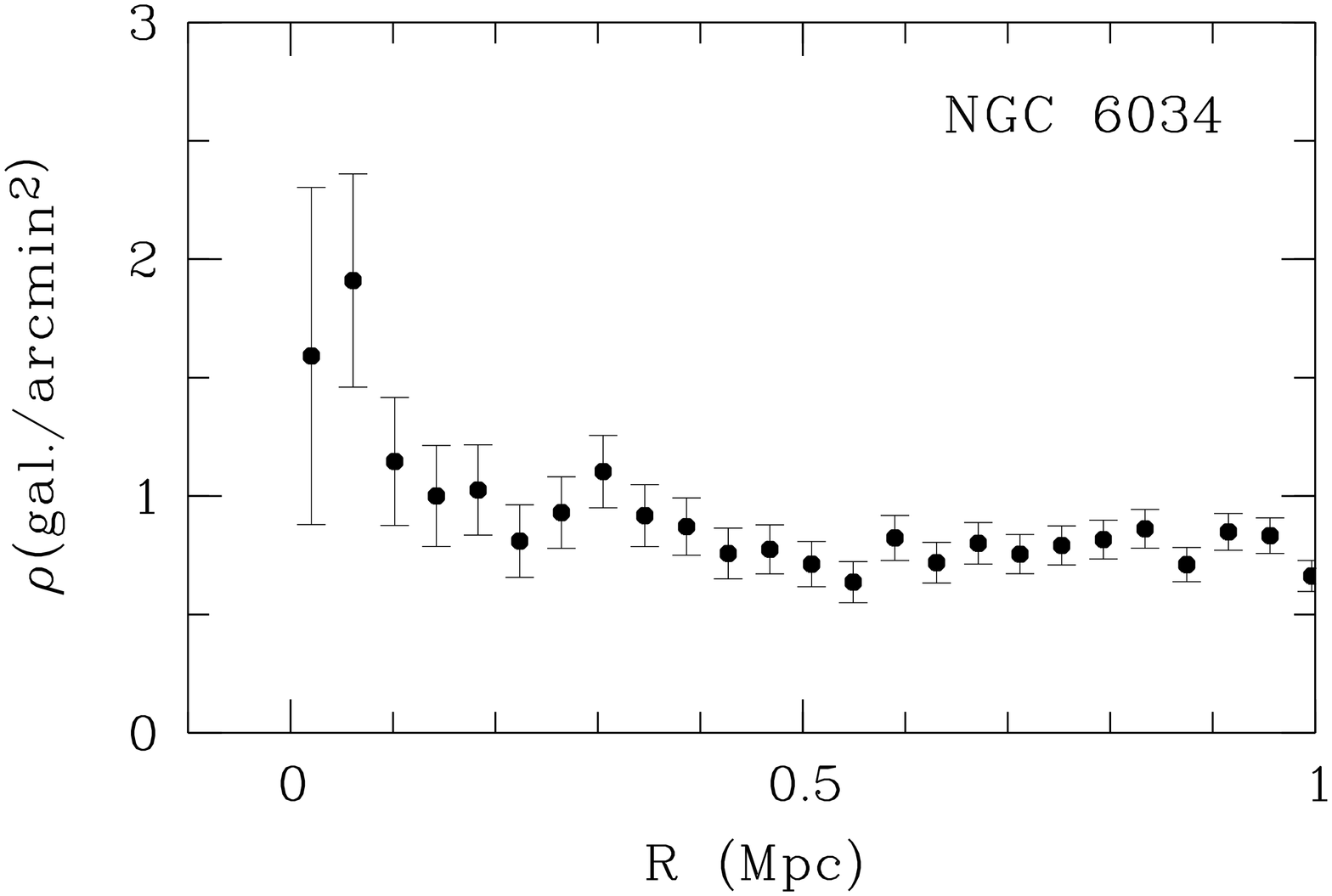}
}
 \caption{Density of galaxies of UGC 842 (left) and NGC 6034 (right) against distance to their central galaxies from SDSS. Only galaxies with magnitudes $<$ 21.5 in the $r$ band and $g-r$ $<$ 1 (color slightly above the red-sequence of galaxies) were considered.}
 \label{fig:dens}
\end{figure*}

From the SDSS DR6 archival we obtain magnitudes and
spectroscopic redshifts of all objects classified as galaxies (class
3 in the SDSS) with $r$'-magnitude $<$ 21 in a 3$\times$3 deg$^{2}$
area around NGC 6034 and in a 2.3$\times$1.7 deg$^{2}$
area around UGC 842. 
The SDSS DR6 catalog in the field of UGC 842 ends at declination $>$ -1.25 deg (J2000), and for 
this reason the analysis performed on SDSS data in this work does not reflect a symmetric field around UGC 842. However, 
our results are not affected by such limitation. The galaxy counts calculated using the objects
classified as galaxies reach their maximum at $r$' $\sim$ 21 mag.
Using this estimate value for the completeness limit and the usual uncertainties in the galaxy classification 
above $r$' $\sim$ 20 mag, we have adopted this latter value as a conservative upper limit for the magnitude. This catalog was then used for the
subsequent analysis.

\section {Analysis and Results}

\subsection{The environment around NGC 6034 and UGC 842}
\label{sct:envngc}

We used the entire SDSS DR 6 galaxy catalog described in Section \ref{sct:sdsscat} to inspect
the environment around UGC 842 and NGC 6034. 
In addition, we used the NASA/IPAC
Extragalactic Database (NED) to search for known groups and clusters of galaxies that could be present in the regions of UGC 842 and NGC 6034. 

The positions of all SDSS objects photometrically classified as galaxies in the fields of UGC 842 and NGC 6034
were used to plot the density of objects ($\rho$) as a function of distance to the
brightest galaxy in each group. These are shown in Figure \ref{fig:dens}. As 
can be noticed, over densities of galaxies are present within circles
of projected radii of about 400 $h^{-1}_{70}$\,kpc 
from the centers of UGC 842 and NGC 6034 (7.4 and 9.8 arcmin, respectively).

\subsubsection{UGC 842}

\begin{figure*}[t!]
\centerline{
\includegraphics[width=80mm]{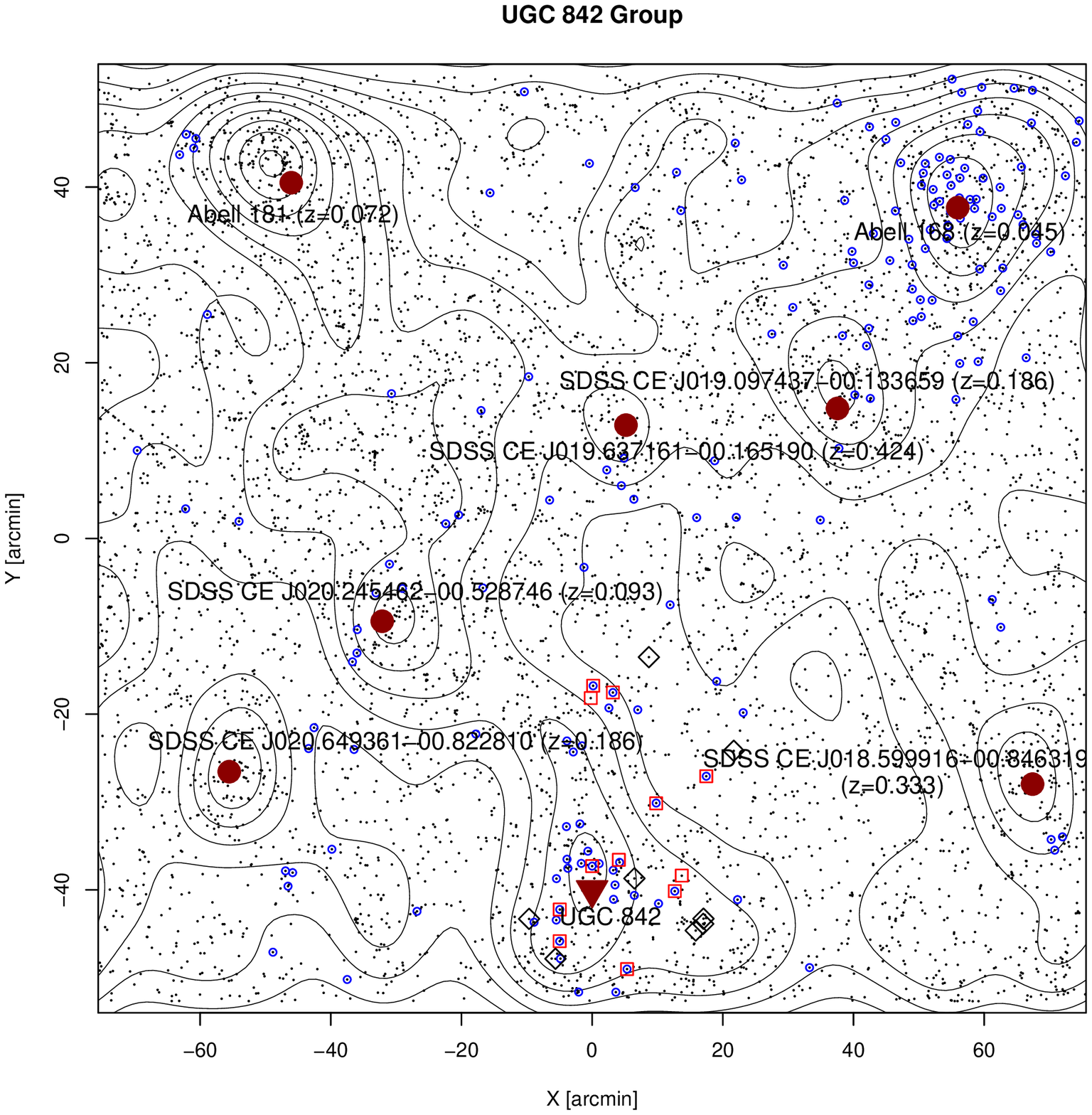}
\includegraphics[width=80mm]{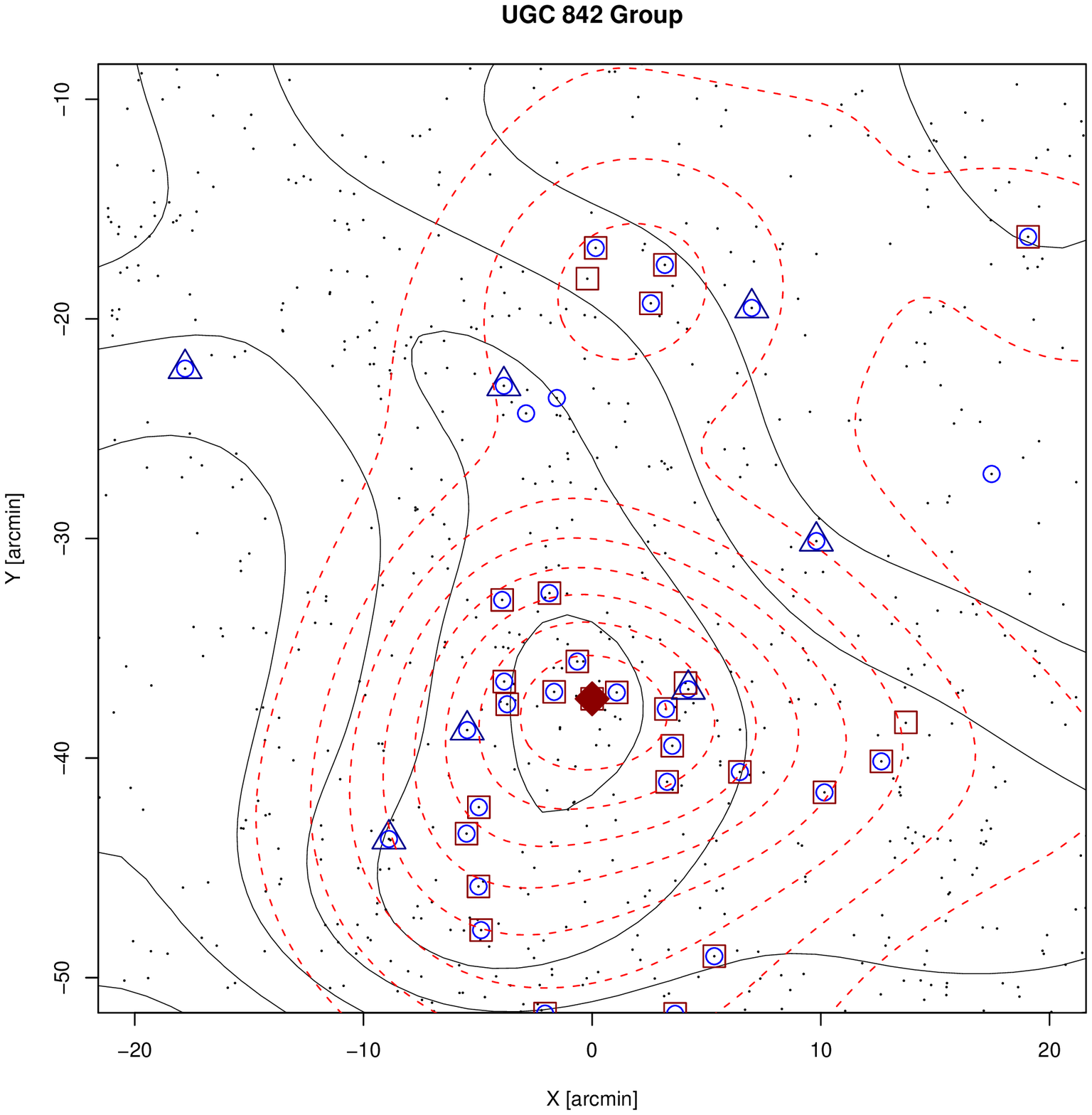}
}
 \caption{
Left: projected galaxy-density map of all galaxies extracted from
the SDSS DR6 with $r$ $<$ 20 mag and within 1.7$\times$1.7 deg$^{2}$ area around UGC 842 ($\sim$
5.4$\times$5.4 Mpc$^{2}$ at the distance 
of the group; gray contours and black dots). The small open
(blue) circles are galaxies identified in the SDSS DR6 catalog with
velocities between 10,000 km\,s$^{-1}$
and 17,000 km\,s$^{-1}$. Galaxies observed with Hydra-CTIO are also plotted.
Open (red) squares: galaxies within the velocity interval of UGC 842;
(black) diamonds: foreground/background galaxies. The big (red) filled circles
indicate the position of the nearby rich clusters Abell 168 ($z$=0.045)
and Abell 181 ($z$=0.072). The filled (red) squares indicate the position of 
clusters detected in the SDSS Commissioning data \citep{Goto02}. The big (red) filled triangle indicates the position of UGC 842
group. 
Right: zoom of the projected galaxy-density map $\sim$ 40$\times$40
 arcmin$^{2}$ ($\sim$ 2.3$\times$2.3 Mpc$^{2}$) around UGC 842. The
(red) dashed contours represent the projected density of galaxies member of
the group. The (red) squares and (blue) triangles represent the passive and star
forming galaxies belonging to the group, respectively.
}
 \label{fig:densmap_UGC}
\end{figure*}

Figure \ref{fig:densmap_UGC}  shows the projected galaxy-density map for the field of UGC 842. It is clear that the 
UGC 842 group is well isolated (at least to the north), with no major or massive galaxy clusters at the same redshift. There are two galaxy clusters that lie closer
in redshift but far in projected distance to the group: Abell 181 at $z$ = 0.072 and Abell 168 at $z$ = 0.045.
Both clusters are located more than 2 deg ($\sim$ 6.4 $h^{-1}_{70}$\,Mpc at $z$ = 0.045) north from UGC 842. Other major
structures are also indicated in the figure, but they all are background clusters.
Figure \ref{fig:densmap_UGC} (right panel) shows a 
zoom of $\sim 80 \times 80$ arcmin$^2$ around the UGC 842 
group. The meaning of the  symbols are the same as in the previous (left panel) figure except 
that we divided the sample of member galaxies in passive (squares) and blue, star forming (triangles). 
The selection of the two samples was performed using the SDSS Database by 
inspecting visually each spectrum.

\begin{figure*}[t!]
\centerline{
\includegraphics[width=80mm]{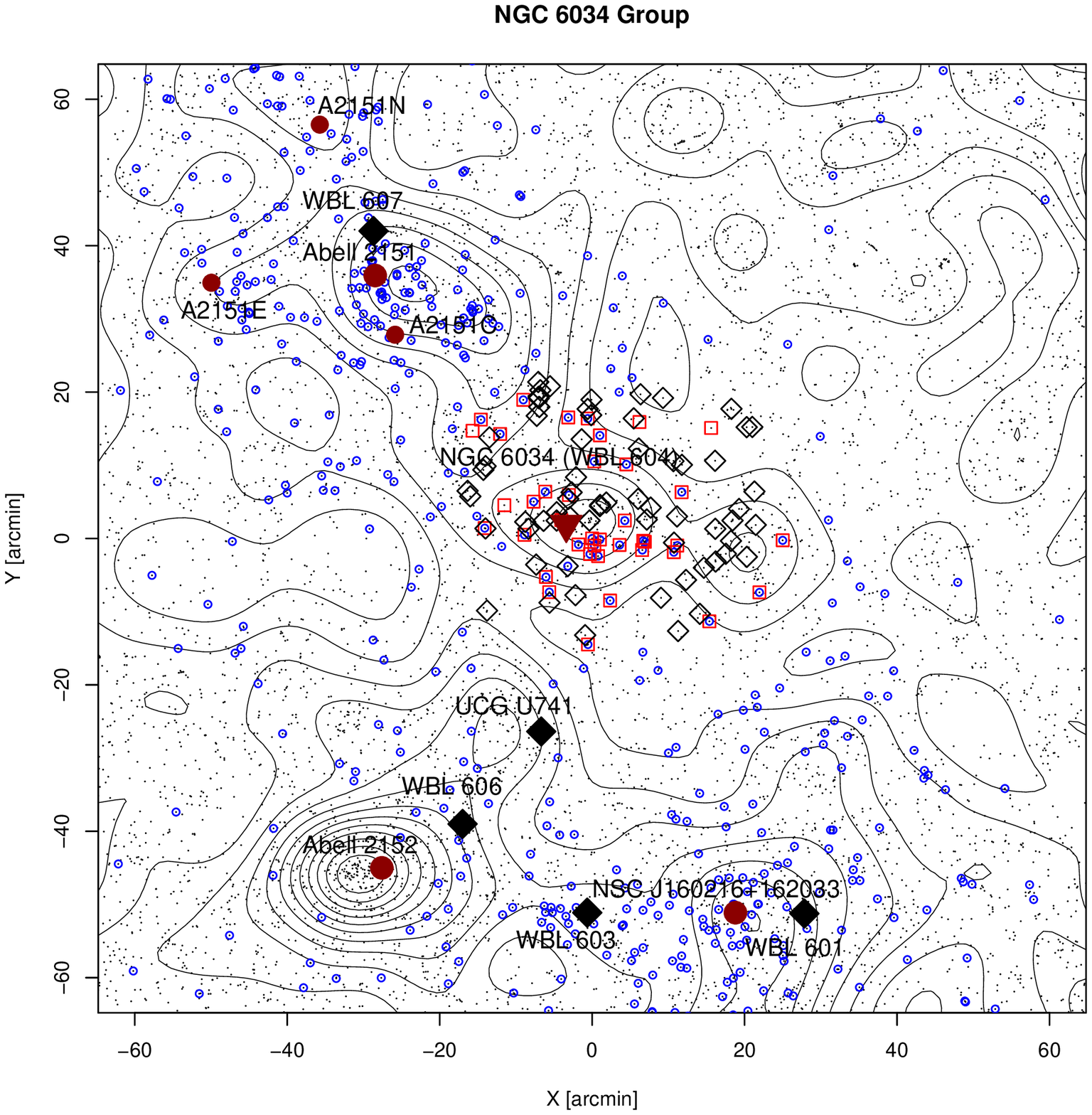}
\includegraphics[width=80mm]{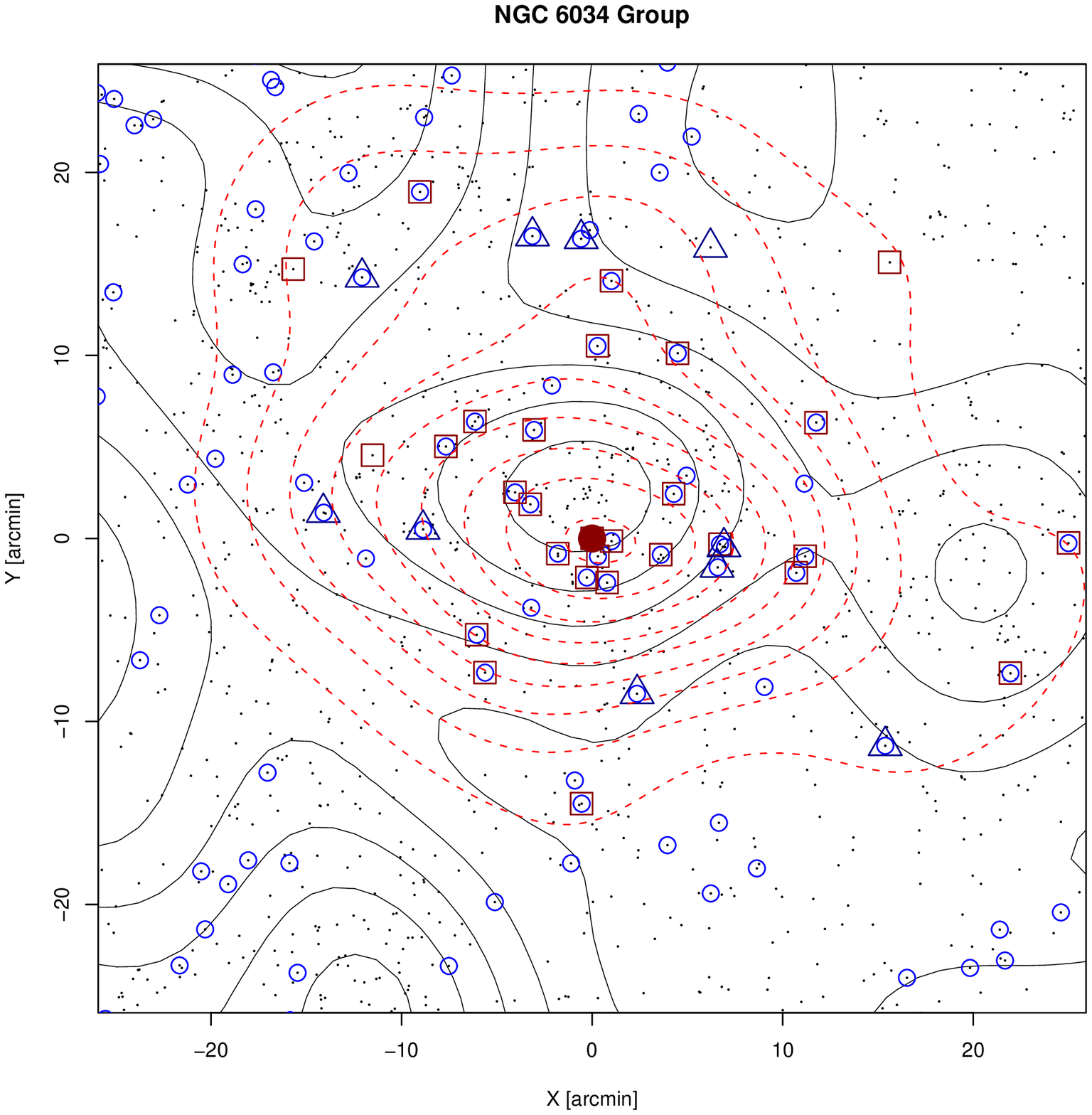}
}
 \caption{
Left: projected galaxy-density map of all galaxies extracted from the SDSS DR6 with
$r$ $<$ 20 mag and within 3 $\times$ 3 deg$^2$ area around NGC 6034 ($\sim$ 7.3$\times$7.3 Mpc$^{2}$ at the distance 
of the group; gray contours and black dots)
. The small (blue) circles are galaxies identified in the SDSS DR6 catalog with velocities
between 9,500 km\,s$^{-1}$ and 12,500 km\,s$^{-1}$. Galaxies observed with Hydra-CTIO are also
plotted. Squares (red): galaxies within the velocity interval of NGC 6034; (black) diamonds:
background galaxies. The big (red) filled circles indicate the position of the major rich clusters,
while the big (black) diamonds show the position of galaxy groups in the area. All inside
the velocity range given above.
Right: zoom of the projected galaxy-density map around NGC 6034 (central 40$\times$40 arcmin$^2$; or $\sim$ 1.6$\times$1.6 Mpc$^{2}$).
The (red) dashed contours
represent the projected density of galaxies with measured velocities in the interval 9,500 km\,s$^{-1}$ and 12,500 km\,s$^{-1}$.
The (red) squares and (blue) triangles represent the passive and star
forming galaxies belonging to the group, respectively.
}
 \label{fig:densmap_NGC}
\end{figure*}

\subsubsection{NGC 6034}

It became clear from our analysis that
NGC 6034 is located in a much richer neighborhood than UGC 842.
Figure \ref{fig:densmap_NGC} (on the left panel) shows the projected 
galaxy-density map of all galaxies with $r$'$<$ 20 mag inside an
area of 3$\times$3 deg$^2$, centered around NGC 6034. In this figure 
one can see that the group is
embedded in a region dominated by three major, rich, galaxy clusters: Abell 2151 (Hercules),
Abell 2152, and NSC J160216+162033 (big filled circles in the figure). Several galaxy groups are
also present in the area: WBL 601, WBL 603, WBL 606, WBL 607, and USGC U741 (big filled diamonds in the figure). Table \ref{tbl:aroundngc}
summarizes the main parameters of the groups and clusters in the region.
Note also that the structure named
NGC 6034 group is not listed in NED and its position corresponds to the
position of the group WBL 604.  

\begin{deluxetable}{llccrl}
\tabletypesize{\scriptsize}
\tablecaption{Groups and clusters around NGC 6034\label{tbl:aroundngc}}
\tablenum{2}
\tablecolumns{6}
\tablewidth{0pc}
\tablehead{
\colhead{Group/Cluster Names} &
\colhead{Type} &
\colhead{R.A.(2000)} &
\colhead{Decl.(2000)} &
\colhead{Velocity} &
\colhead{Velocity Reference}\\
\colhead{} &
\colhead{} &
\colhead{} &
\colhead{} &
\colhead{(km s$^{-1}$)} &
\colhead{}}
\startdata
Abell 2151         & Cluster & 16:05:25.9 & $+$17:47:50 & 11070 & \citet{Bird95}    \\
Abell 2151E        & Cluster & 16:06:51.9 & $+$17:46:51 &  9623 & \citet{Ebeling98} \\
Abell 2151C        & Cluster & 16:05:15.5 & $+$17:39:45 & 10650 & \citet{Bird95}    \\
Abell 2151N        & Cluster & 16:05:55.0 & $+$18:08:27 & 11445 & \citet{Bird95}    \\
Abell 2152         & Cluster & 16:05:22.4 & $+$16:26:55 & 12291 & \citet{Struble99} \\
NSC J160216+162033 & Cluster & 16:02:16.9 & $+$16:20:32 & 11422 & \citet{Gal03}     \\
WBL 601            & Group   & 16:01:40.7 & $+$16:20:41 & 10637 & \citet{White99}   \\
WBL 603            & Group   & 16:03:34.6 & $+$16:20:48 & 11212 & \citet{White99}   \\
WBL 606            & Group   & 16:04:40.4 & $+$16:32:52 &  9539 & \citet{White99}   \\
WBL 607            & Group   & 16:05:26.8 & $+$17:53:55 & 10936 & \citet{White99}   \\
UCG U741           & Group   & 16:03:58.8 & $+$16:45:33 & 10821 & \citet{Ramella00} \\
\enddata
\end{deluxetable}

\begin{figure*}[t!]
\centerline{
\includegraphics[angle=-90,width=70mm]{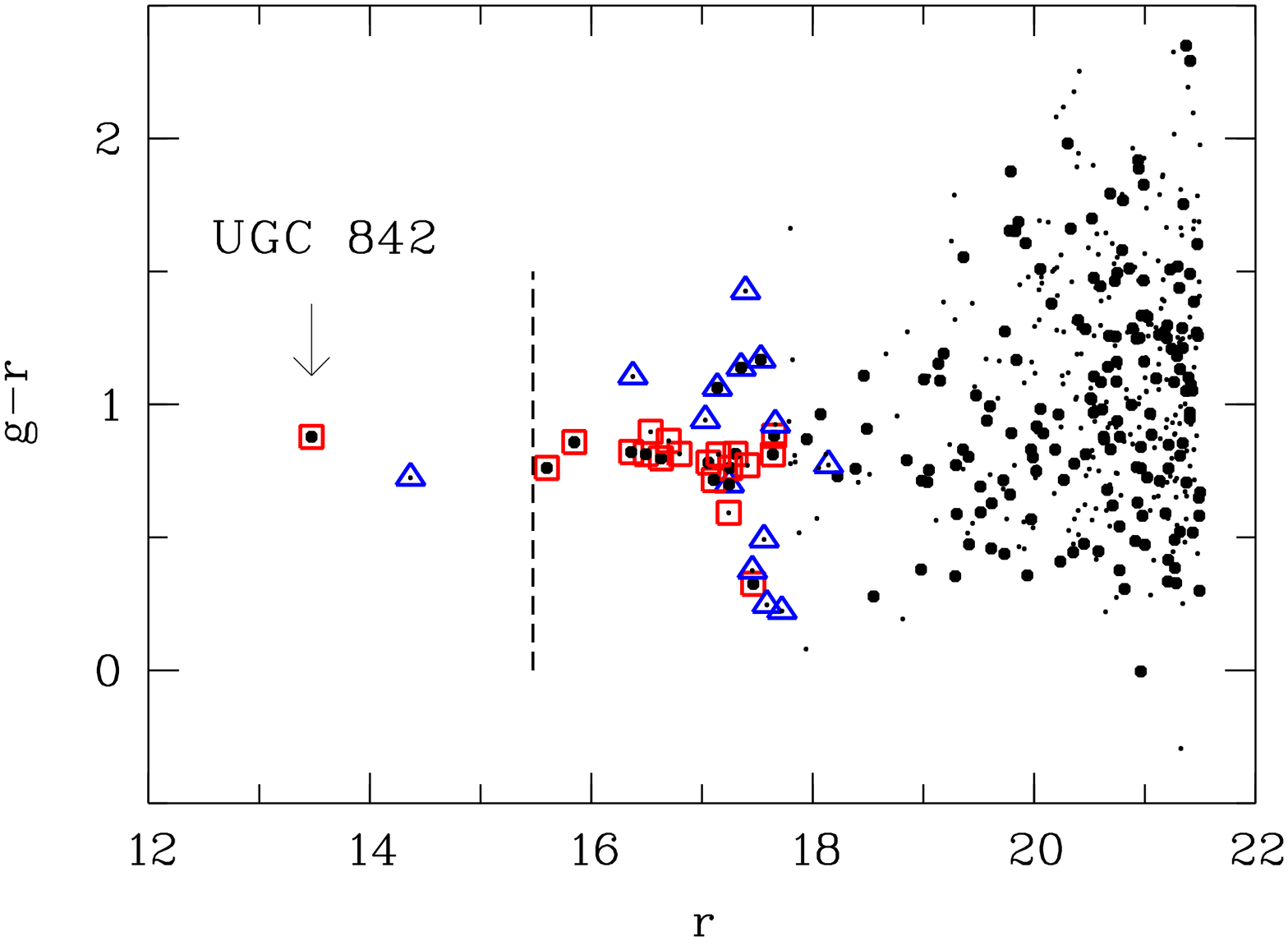}~~~~~~~~~~
\includegraphics[angle=-90,width=70mm]{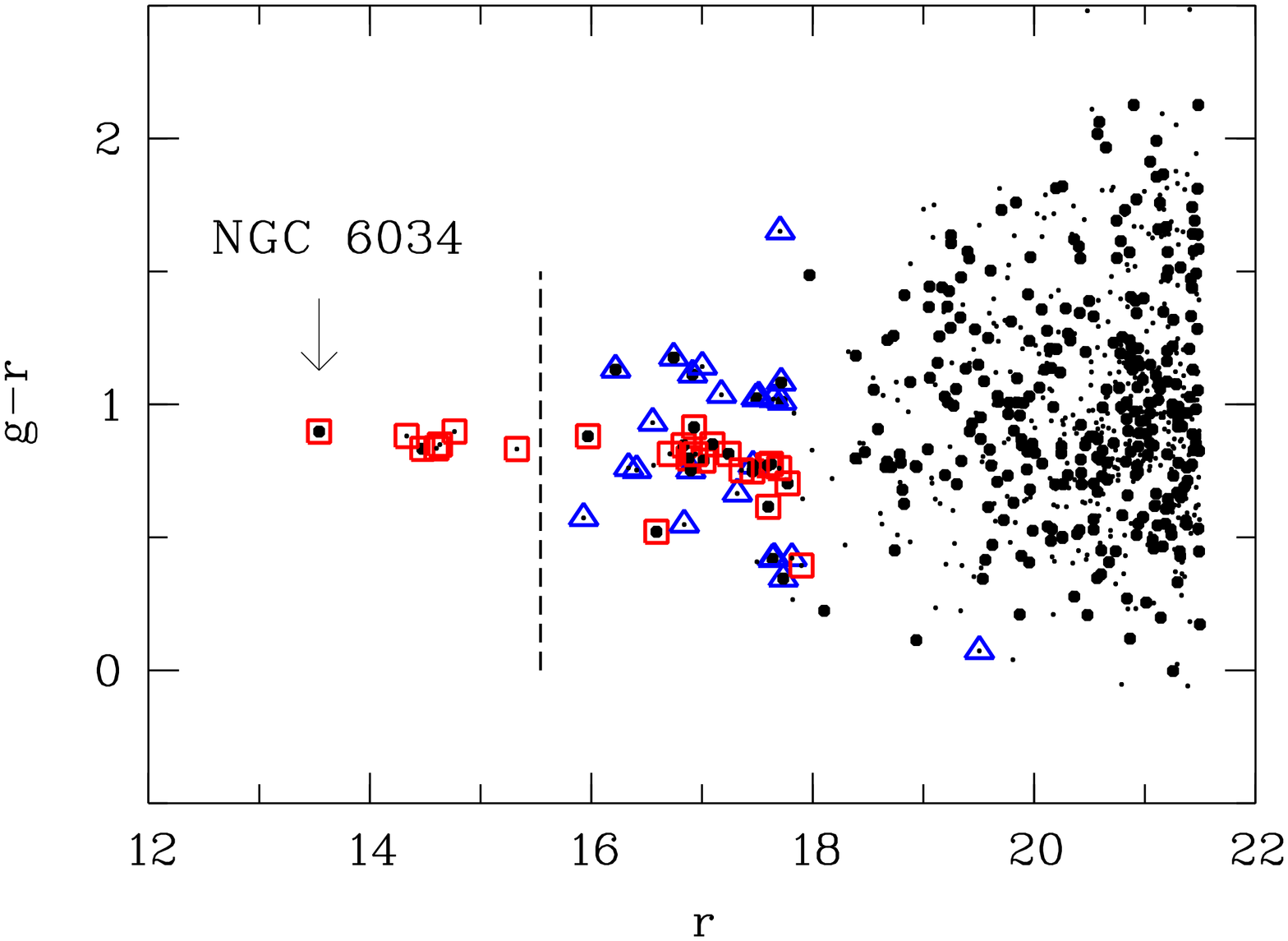}
}
 \caption{Color-magnitude diagrams of UGC 842 (left) and NGC 6034, using spectroscopic information from SDSS and our own spectroscopy. The big dots represent galaxies within a circle with radius of 300 kpc centered at the brightest galaxy of each group, 
while small dots correspond to galaxies inside an annulus between 300 and 500 kpc. The group members are marked by boxes and non-members by triangles. The dashed lines show the magnitudes in the $r$ band corresponding to a $\Delta r$ = 2 with respect to the brightest galaxies.}
 \label{fig:cmd}
\end{figure*}

In order to investigate how many galaxy groups belong to the large
structure in which NGC 6034 seems to be embedded we show in Figure
\ref{fig:densmap_NGC} (on the right side) a zoom of the 40$\times$40 arcmin$^2$
field around this group. 
From the figure it is clear that 
a number of galaxies with velocities in the interval 9,000 and 12,000
 km\,s$^{-1}$ may in fact belong to other
nearby structures. The (red) dashed contours
in the figure represent the projected density of galaxies at the
distance of the group. We can see that the peak of the distribution
is offset by 3'--4' from the high density peak given by the projected
density map obtained using all galaxies with $r$' $<$ 20 mag. The
reason of this displacement is due to the existence of two massive
structures along the line of sight at $z$ $\sim$ 0.11 and $z$ $\sim$
0.14 (see Figure \ref{fig:veldist}, left).

\subsection{Color-Magnitude Diagrams \label{sct:cmd}}

A red-sequence of galaxies is clearly seen in the color-magnitude
diagram (CMD) of both groups (Figure \ref{fig:cmd}). 
Group membership is confirmed by
spectroscopy for all galaxies with $r$ $<$ 17.7 mag inside 500 $h_{70}^{-1}$\,kpc around UGC 842 and NGC 6034, with two exceptions for NGC 6034.

The UGC 842 group obeys the optical criteria to be classified as a fossil group,
in which the brightest galaxy dominates the group within R$_{vir}$/2.
The two closest ``bright'' galaxies are yet relatively far from UGC 842 (J011913.49-010839.9: $d_{proj}$ $\sim$ 528 $h_{70}^{-1}$\,kpc, $r$ = 14.58 mag; and J011832.25-011150.4: $d_{proj}$
$\sim$ 688 $h_{70}^{-1}$\,kpc, $r$ = 14.66 mag).
The same is not true for the NGC 6034 group.
Six galaxies in the field of NGC 6034, shown in
Figure \ref{fig:cmd}, violate the magnitude criteria adopted to
classify a group as a fossil group within R$_{vir}$/2 
(see parameters in Table \ref{tab:vel}).
One of them,
J160348.24+171426.2, has a $\Delta r$ $\sim$ 0.9 mag, $d_{proj}$
$\sim$ 188 $h_{70}^{-1}$\,kpc, and $\Delta$$V$ $\sim$ 800 km\,s$^{-1}$, with respect to the
brightest galaxy. It is not completely clear if this galaxy, with
such a relatively high $\Delta$$V$ is or is not a group member -- note that $\sigma_v$ for this group is $\sim$ 230 km\,s$^{-1}$. The
other five identified members which also break the fossil group definition have
300 $h_{70}^{-1}$\,kpc $<$ $d_{proj}$ $<$ 460 $h_{70}^{-1}$\,kpc, and $\Delta$V $<$ 240 km\,s$^{-1}$.
They are: 
J160402.75+171656.6,
J160419.55+171049.4,
J160249.22+171002.6,
J160314.11+172202.4, and
J160356.65+171818.4.

\subsection{SDSS DR6 Versus CTIO Observations}
\label{sct:sdsshydra}

\begin{figure}[t!]
\centerline{
\includegraphics[bb=0cm 0cm 20cm 20cm,clip=true,width=75mm]{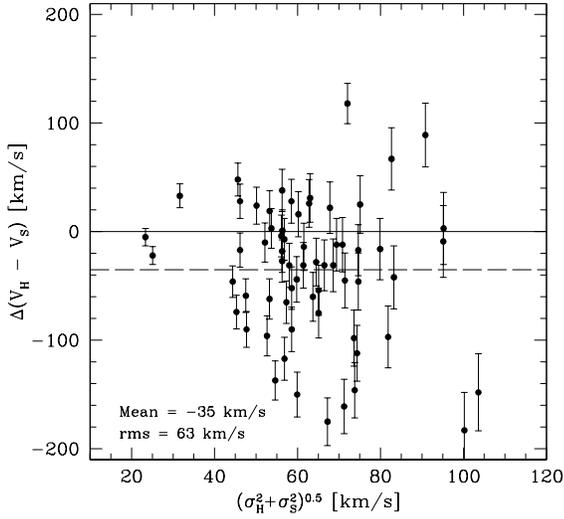}
}
 \caption{
Residual of the heliocentric velocity as a function of the internal quadratic 
errors for 64 galaxies (NGC 6034 and UGC 842) in common with SDSS DR6. The 
mean shift of the data is 35 km\,s$^{-1}$ with a {\it rms} of 63 km\,s$^{-1}$.
}
 \label{fig:resvel}
\end{figure}

From the initial SDSS catalog described in Section \ref{sct:sdsscat}, 
we concentrated on all galaxies inside a region of radius $<$ 25'
around UGC 842 and NGC 6034 (the region covered by our Hydra-CTIO observations)
and with velocities between 9,000 km\,s$^{-1}$ and 60,000 km\,s$^{-1}$.
In this velocity interval, we have 55 and 9 galaxies that were observed
in common with SDSS DR6 and our observations in the area of UGC 842 and NGC 6034, respectively. 
We used this information 
to search for any possible systematic effects and  
to estimate the true errors in our measurements. 
Figure \ref{fig:resvel} shows the
residuals of the velocities (our velocities, V$_{H}$, minus velocities from the SDSS catalog, V$_S$) 
as a function of the quadratic errors
($\sigma_H$ and $\sigma$$_S$)
for the 64 galaxies in common with the SDSS DR6 data set. The average
difference between the two data sets is 35 km\,s$^{-1}$ with and
$rms$ of 63 km\,s$^{-1}$. 
Thus, in order to minimize the errors in the determination
of the average velocity and the velocity dispersion, we applied
this small velocity correction to all galaxies in the SDSS DR6
catalog. 

For NGC 6034, we selected all galaxies with radial velocities in 
the interval 9,000--12,000 km\,s$^{-1}$. A total of 35 galaxies are in
common with the SDSS DR6. For five of the galaxies observed with
Hydra-CTIO, the velocities were obtained for the first time. In
addition, there are six galaxies listed in the SDSS DR6 and inside
the above interval that were not observed by us. In total, we have
46 galaxies in the area of NGC 6034 and within the inverval of radial velocities of 
9,000 to 12,000 km\,s$^{-1}$.

For UGC 842, 
we selected all galaxies with radial velocities in 
the interval 10,000--17,000 km\,s$^{-1}$. A total of nine galaxies are in
common with the SDSS DR6. For three of the galaxies observed with
Hydra-CTIO, the velocities were obtained for the first time.

\subsection{Velocity distributions and mass determinations}
\label{sct:veldist}

The new galaxy radial velocities in the fields of UGC 842 and NGC 6034 are listed in Tables \ref{tbl:ugcvel} and \ref{tbl:ngcvel}, respectively. 
The radial velocities and the associated errors derived using cross-correlation (absorption-line systems) are given in the column named V$^{\rm CC}_{\rm hel}$.
The radial velocities from emission lines are given in the column named V$^{\rm EM}_{\rm hel}$. All radial velocities are corrected 
to the heliocentric reference frame. The galaxy identifications and their coordinates, total magnitudes in the $g$' band and $r$' band, $R$ values, and the number of emission lines ($\#$el) used when it is 
possible for velocity calculation (see Section \ref{sct:rv}) are also given in the tables. 
We used the robust bi-weight estimators CBI and SBI of \citet{Beers90} to calculate a reliable value 
for the average velocity and the velocity dispersion for each group. 
The adopted radial velocities
in Tables \ref{tbl:ugcvel} and \ref{tbl:ngcvel} for the determination of the average velocity and 
velocity dispersion of the groups are always the velocities calculated by cross-correlation, 
except in the cases when the S/N of the absorption line spectra was too low (in one case for the UGC 842 group and five cases for the NGC 6034 group).
We use the relation of \citet{Carlberg97} in order to estimate the virial radius of the systems: 
$r_{vir}$ = $\sqrt{3}$/10$\times$ $\sigma_v$/H($z$), 
where H($z$)=H$_0 \times$[$\Omega_M$(1+$z$)$^3$+(1-$\Omega_M$-$\Omega_{\Lambda}$)(1+$z$)$^{2}$+$\Omega_{\Lambda}$]$^{1/2}$.
Then the mass was estimated by $M_{virial}$ = (2/G)$\sigma_v^2$$r_{virial}$.

\clearpage
\begin{deluxetable}{rrcccccrrrr}
\tabletypesize{\scriptsize}
\tablecaption{New galaxy Radial Velocities in the Field of UGC 842\label{tbl:ugcvel}}
\tablenum{3}
\tablecolumns{11}
\tablewidth{0pc}
\tablehead{
\colhead{Gal.} &
\colhead{Fib.} &
\colhead{SDSS ID} &
\colhead{$g^{'}_{SDSS}$}&
\colhead{$r^{'}_{SDSS}$}&
\colhead{$r$}&
\colhead{$b$}&
\colhead{$V_{\rm hel}^{\rm CC}$} &
\colhead{$R$} &
\colhead{V$_{\rm hel}^{\rm EM}$} &
\colhead{\#el} \\
\colhead{} &
\colhead{} &
\colhead{} &
\colhead{(mag)}&
\colhead{(mag)}&
\colhead{(mag)}&
\colhead{(mag)}&
\colhead{(km s$^{-1}$)} &
\colhead{} &
\colhead{(km s$^{-1}$)} &
\colhead{} 
}
\startdata

  248 &  66 & J011726.91-004700.1 & 18.81 & 18.41 & 18.08 & 19.10 & 17339$\pm$69 &   2.42 & \nodata    & \nodata \\
 1309 &  21 & J011743.74-004953.0 & 16.00 & 15.18 & 14.93 & 16.48 & 15930$\pm$35 &  10.07 & \nodata    & \nodata \\
 3134 &   5 & J011745.47-010606.2 & 19.02 & 18.54 & 18.24 & 19.38 & 44194$\pm$84 &   2.58 & \nodata    & \nodata \\
 3079 &  79 & J011745.62-010641.5 & 19.66 & 19.04 & 18.77 & 20.07 &  6928$\pm$113&   2.62 & \nodata    & \nodata \\
 3755 & 114 & J011750.18-010727.2 & 19.60 & 18.78 & 18.88 & 20.35 & 51098$\pm$82 &   4.14 & \nodata    & \nodata \\
 4780 &  54 & J011758.75-010112.8 & 18.76 & 18.47 & 18.20 & 19.03 & 14190$\pm$101&   3.06 & \nodata    & \nodata \\
 5256 &  11 & J011803.03-010258.1 & 17.12 & 16.21 & 16.03 & 17.68 & 13192$\pm$67 &   5.24 & \nodata    & \nodata \\
 7155 &  52 & J011814.40-005256.3 & 16.85 & 16.43 & 16.13 & 17.20 & \nodata             & \nodata&14781$\pm$85&    6  \\
 8131 &  67 & J011818.87-003619.6 & 19.53 & 19.05 &\nodata&\nodata& 26548$\pm$60 &   4.93 & \nodata    & \nodata \\
 9106 &  12 & J011827.75-010131.4 & 17.83 & 17.59 & 17.24 & 18.18 &  \nodata             & \nodata& 5868$\pm$72&   6  \\
 7894 & 108 & J011832.25-011150.4 & 15.61 & 14.66 & 14.39 & 16.07 & 13873$\pm$89 &  4.20  & \nodata    & \nodata  \\
 1002 &  43 & J011837.30-005924.2 & 16.70 & 15.85 & 15.60 & 17.19 & 13987$\pm$66 &  5.56  & \nodata    & \nodata  \\
 1126 &  15 & J011840.90-004021.7 & 17.44 & 16.62 & 16.37 & 17.92 & 14627$\pm$59 &  6.76  & \nodata    & \nodata  \\
 1187 &  18 & J011852.99-003935.3 & 16.08 & 15.20 & 14.94 & 16.58 & 14311$\pm$33 &  2.07  & \nodata    & \nodata  \\
 8901 &  39 & J011853.62-010007.2 & 14.35 & 13.47 & 13.21 & 14.85 & 13428$\pm$40 & 10.09  & \nodata    & \nodata  \\
13007 &  37 & J011854.47-004059.5 & 15.93 & 15.04 & 14.78 & 16.44 & 13511$\pm$35 & 11.30  & \nodata    & \nodata  \\
 1597 &  93 & J011913.43-010503.4 & 17.56 & 16.70 & 16.47 & 18.14 & 13771$\pm$58 &  4.94  & \nodata    & \nodata  \\
 1480 & 124 & J011913.49-010839.9 & 15.55 & 14.58 & 14.30 & 16.08 & 13789$\pm$44 &  8.70  & \nodata    & \nodata  \\
 1662 &  31 & J011916.05-011040.2 & 19.17 & 18.81 & 18.49 & 19.53 &  8358$\pm$77 &  3.89  & \nodata    & \nodata  \\
 1821 & 128 & J011932.22-010607.9 & 18.16 & 17.39 & 16.94 & 18.48 & 61195$\pm$42 &  4.34  & \nodata    & \nodata  \\
											   
\enddata
\end{deluxetable}
 
 \clearpage

\begin{deluxetable}{rrccccccrrrr}
\tabletypesize{\scriptsize}
\tablecaption{New Galaxy Radial Velocities in the Field of NGC 6034 \label{tbl:ngcvel}}
\tablenum{4}
\tablecolumns{12}
\tablewidth{0pc}
\tablehead{
\colhead{Gal.} &
\colhead{Fib.} &
\colhead{Run} &
\colhead{SDSS ID} &
\colhead{$g^{'}_{SDSS}$}&
\colhead{$r^{'}_{SDSS}$}&
\colhead{$r$}&
\colhead{$b$}&
\colhead{V$_{\rm hel}^{\rm CC}$} &
\colhead{$R$} &
\colhead{V$_{\rm hel}^{\rm EM}$} &
\colhead{\#el} \\
\colhead{} &
\colhead{} &
\colhead{} &
\colhead{} &
\colhead{(mag)}&
\colhead{(mag)}&
\colhead{(mag)}&
\colhead{(mag)}&
\colhead{(km s$^{-1}$)} &
\colhead{} &
\colhead{(km s$^{-1}$)} &
\colhead{} 
}
\startdata

     1 &   87 & 1 & J160152.09+171139.8 & 17.83 & 16.96 &  ...  &  ...  & 10499$\pm$33 & 10.68 & \nodata      & \nodata \\
   333 & 63 & 1 & J160204.26+170433.0 & 14.47 & 13.60 &  ...  &  ...  & 10957$\pm$16 & 14.68 & \nodata      & \nodata \\
   780 & 129 & 1 & J160206.23+171345.4 & 18.08 & 16.99 &  ...  &  ...  & 42264$\pm$66 &  8.42 & \nodata      & \nodata \\
   857 & 137 & 1 & J160206.90+171819.8 & 18.80 & 18.38 &  ...  &  ...  & 49763$\pm$61 &  3.98 & \nodata      & \nodata \\
   885 & 72 & 1 & J160208.02+172708.4 & 18.62 & 17.50 & 17.35 & 19.29 & 42339$\pm$38 &  9.10 & \nodata      & \nodata \\
  1042 & 104 & 1 & J160210.90+170925.4 & 19.45 & 18.00 & 16.82 & 19.14 & 69098$\pm$72 &  5.56 & \nodata      & \nodata \\
  1076 & 130 & 1 & J160210.98+172707.1 & 18.00 & 17.25 & 17.05 & 18.48 & 32522$\pm$43 &  8.26 & \nodata      & \nodata \\
  1278 & 49 & 2 & J160214.96+171557.4 & 18.49 & 17.88 & 17.94 & 19.22 & \nodata      &\nodata& 42010$\pm$23 &     5	\\
  1363 & 48 & 1 & J160218.39+171421.3 & 17.84 & 16.72 & 16.54 & 18.56 & 42384$\pm$48 &  9.12 & \nodata      & \nodata \\
  1473 & 115 & 1 & J160219.00+172938.4 & 20.51 & 18.78 & 17.65 & 20.21 & 92751$\pm$87 &  7.40 & \nodata      & \nodata \\
  1438 & 132 & 1 & J160219.24+171134.7 & 19.25 & 18.13 & 17.91 & 20.01 & 42523$\pm$53 &  7.60 & \nodata      & \nodata \\
  1426 & 19 & 1 & J160221.89+170949.3 & 18.65 & 17.52 &  ...  &  ...  & 42061$\pm$56 &  8.23 & \nodata      & \nodata \\
  1830 & 81 & 2 & J160227.29+170845.9 & 18.27 & 17.17 & 16.85 & 18.80 & 42394$\pm$68 &  6.64 & \nodata      & \nodata \\
  1873 & 8 & 2 & J160227.48+171313.6 & 18.35 & 17.51 &  ...  &  ...  & 42429$\pm$77 &  2.80 & \nodata      & \nodata \\
  1888 & 111 & 1 & J160227.52+172232.3 & 18.78 & 18.06 & 17.80 & 19.23 & 41834$\pm$48 &  5.12 & \nodata      & \nodata \\
  1908 & 23 & 1 & J160229.59+172700.3 & 19.11 & 18.35 & 18.06 & 19.51 & 11034$\pm$53 &  5.81 & \nodata      & \nodata \\
  1994 & 6 & 1 & J160230.54+170035.6 & 17.82 & 17.06 & 16.81 & 18.24 &  9842$\pm$31 &  9.58 & \nodata      & \nodata \\
\nodata& 114 & 2 & \nodata		&\nodata&\nodata&\nodata&\nodata&  9808$\pm$41 &  8.98 & 9922$\pm$49  &    10	\\
  2144 & 82 & 1 & J160233.45+170754.4 & 18.57 & 18.11 & 17.82 & 18.93 & \nodata      &\nodata&40973$\pm$12  &     5	\\
  2265 & 100 & 1 & J160235.55+170136.9 & 19.02 & 18.24 & 17.97 & 19.48 & 36300$\pm$96 &  2.40 & \nodata      & \nodata \\
  2588 & 57 & 1 & J160242.80+170614.3 & 19.19 & 17.92 & 17.71 & 19.98 & 52668$\pm$40 &  7.59 & \nodata      & \nodata \\
  2677 & 92 & 1 & J160244.90+172159.4 & 18.62 & 18.17 & 17.97 & 19.07 & 18319$\pm$80 &  3.44 & \nodata      & \nodata \\
  2580 & 66 & 1 & J160245.04+171815.4 & 16.45 & 15.82 & 15.58 & 16.83 & 10678$\pm$51 &  6.90 & \nodata      & \nodata \\
  2720 & 12 & 1 & J160246.95+165916.5 & 19.30 & 18.30 & 17.85 & 19.67 & 50271$\pm$65 &  5.97 & \nodata      & \nodata \\
  2746 & 22 & 2 & J160247.36+171056.6 & 17.52 & 16.71 & 16.77 & 18.30 & 10282$\pm$24 & 12.84 & \nodata      & \nodata \\
  2716 & 8 & 1 & J160247.54+171454.9 & 17.16 & 16.41 &  ...  &  ...  & 13051$\pm$48 &  7.01 & 13160$\pm$63 &     5	\\
  2801 & 54 & 1 & J160249.11+171120.4 & 18.21 & 17.17 & 16.87 & 18.65 & 32196$\pm$46 &  6.03 & \nodata      & \nodata \\
  2691 & 79 & 1 & J160249.22+171002.6 & 15.48 & 14.63 & 14.41 & 16.01 & 10356$\pm$27 & 19.49 & \nodata      & \nodata \\
  2858 & 103 & 1 & J160249.29+172228.4 & 20.89 & 20.53 & 17.62 & 18.61 & 15987$\pm$100&  3.53 & \nodata      & \nodata \\
  3033 & 61 & 1 & J160254.86+173106.4 & 17.74 & 16.71 & 16.59 & 18.47 & 30006$\pm$23 & 13.37 & \nodata      & \nodata \\
  3158 & 116 & 1 & J160255.92+170348.3 & 18.08 & 17.65 & 17.35 & 18.42 & \nodata      &\nodata& 13810$\pm$10 &     6	\\
  3344 & 69 & 1 & J160301.29+171609.5 & 17.98 & 17.32 & 17.10 & 18.46 & 33203$\pm$77 &  3.20 & \nodata      & \nodata \\
  3524 & 46 & 1 & J160303.39+171429.1 & 18.73 & 17.72 & 17.47 & 19.24 & 40356$\pm$34 &  8.86 & \nodata      & \nodata \\
\nodata& 85 & 2 & \nodata		&\nodata&\nodata&\nodata&\nodata& 40335$\pm$88 &  3.99 & \nodata      & \nodata \\
  3615 & 25 & 1 & J160305.24+171136.1 & 17.67 & 16.88 & 16.74 & 18.32 &  9952$\pm$38 & 10.62 & \nodata      & \nodata \\
  3379 & 86 & 1 & J160305.72+171020.3 & 17.11 & 16.59 & 16.26 & 17.36 & \nodata      &\nodata& 10050$\pm$40 &     7   \\
  3651 & 84 & 1 & J160306.68+173136.4 & 17.61 & 16.95 & 16.88 & 18.05 & 29801$\pm$23 & 10.56 & \nodata      & \nodata \\
  3686 & 33 & 1 & J160307.22+172749.4 & 17.36 & 16.88 & 16.61 & 17.69 & 10776$\pm$51 &  4.35 & 10835$\pm$36 &     9   \\
  3580 & 67 & 1 & J160307.58+172412.9 & 18.39 & 17.87 & 17.69 & 18.77 & 33031$\pm$82 &  3.41 & 32942$\pm$40 &     6   \\
  3910 & 68 & 1 & J160307.88+171719.9 & 18.67 & 17.65 & 17.35 & 19.07 & 40339$\pm$57 &  5.36 & \nodata      & \nodata \\
  3998 & 16 & 1 & J160310.23+172819.0 & 18.25 & 17.58 & 17.38 & 18.68 & 33765$\pm$38 &  5.99 & \nodata      & \nodata \\
  4168 & 15 & 1 & J160314.11+172202.4 & 15.66 & 14.76 & 14.49 & 16.18 &  9943$\pm$41 & 12.47 & \nodata      & \nodata \\
  4221 & 90 & 1 & J160314.95+171421.3 & 18.20 & 17.45 & 17.05 & 18.51 & 10487$\pm$37 &  7.84 & \nodata      & \nodata \\
  4411 & 108 & 1 & J160317.65+171101.7 & 17.68 & 16.83 & 16.76 & 18.32 & 10435$\pm$53 &  5.77 & \nodata      & \nodata \\
  4527 & 99 & 1 & J160322.65+170326.0 & 17.10 & 16.34 &  ...  &  ...  & 11783$\pm$25 & 12.06 & \nodata      & \nodata \\
\nodata& 45 & 2 & \nodata		&\nodata&\nodata&\nodata&\nodata& 11805$\pm$48 & 11.39 & 11717$\pm$52 &    10   \\
  4729 & 42 & 1 & J160324.62+171648.8 & 17.92 & 16.74 & 16.40 & 18.40 & 40197$\pm$30 & 13.00 & \nodata      & \nodata \\
  4930 & 59 & 1 & J160327.88+171628.7 & 20.16 & 19.06 & 17.57 & 19.40 & 40581$\pm$95 &  3.60 & \nodata      & \nodata \\
  4920 & 124 & 1 & J160328.00+171146.2 & 17.84 & 16.93 & 16.76 & 18.38 & 10314$\pm$34 & 12.36 & \nodata      & \nodata \\
\nodata& 29 & 2 & \nodata		&\nodata&\nodata&\nodata&\nodata& 10339$\pm$35 & 11.63 & \nodata      & \nodata \\
  4913 & 91 & 2 & J160328.00+171619.8 & 18.03 & 16.91 & 16.89 & 18.83 & 40418$\pm$79 &  5.79 & \nodata      & \nodata \\
  4749 & 91 & 1 & J160328.02+172559.4 & 16.09 & 15.24 & 15.03 & 16.65 & 10568$\pm$22 & 24.08 & \nodata      & \nodata \\
  4989 & 31 & 1 & J160328.93+170930.2 & 18.40 & 17.62 & 17.58 & 19.06 & 10563$\pm$32 &  8.74 & \nodata      & \nodata \\
\nodata& 14 & 2 & \nodata		&\nodata&\nodata&\nodata&\nodata& 10638$\pm$74 &  4.19 & \nodata      & \nodata \\
  5099 & 93 & 2 & J160330.86+171056.4 & 18.05 & 17.24 & 16.96 & 18.41 & 10003$\pm$32 & 12.22 & \nodata      & \nodata \\
  5065 & 20 & 1 & J160330.94+172226.0 & 17.75 & 16.94 & 16.58 & 18.11 & 10307$\pm$28 &  9.30 & \nodata      & \nodata \\
  4090 & 29 & 1 & J160332.08+171155.2 & 14.44 & 13.54 & 13.08 & 14.69 & 10162$\pm$48 & 10.75 & \nodata      & \nodata \\
  5204 & 18 & 1 & J160332.41+173051.9 & 18.53 & 17.65 &\nodata&\nodata& 29715$\pm$21 & 18.43 & \nodata      & \nodata \\
  4972 & 37 & 1 & J160332.58+172845.9 & 15.70 & 15.01 & 14.78 & 16.06 & 13197$\pm$22 & 14.83 & \nodata      & \nodata \\
  5232 & 9 & 2 & J160333.18+170947.1 & 17.80 & 17.01 & 16.95 & 18.43 &  9985$\pm$31 & 12.02 & \nodata      & \nodata \\
  4846 & 64 & 1 & J160333.42+171421.6 & 17.35 & 16.22 & 15.45 & 17.47 & 40951$\pm$45 & 10.05 & \nodata      & \nodata \\
  5091 & 135 & 1 & J160334.27+165725.7 & 16.26 & 15.48 & 15.06 & 16.49 & 10064$\pm$31 & 13.64 & \nodata      & \nodata \\
  5269 & 117 & 1 & J160334.38+172817.5 & 17.76 & 17.46 & 16.98 & 17.95 & \nodata      &\nodata& 11384$\pm$71 &      6  \\
  5306 & 34 & 2 & J160334.45+172936.3 & 19.57 & 18.11 & 17.74 & 20.15 & 67955$\pm$80 &  5.33 & \nodata      & \nodata \\
  5348 & 9 & 1 & J160335.72+165841.7 & 17.87 & 17.22 & 16.86 & 18.41 & 14316$\pm$51 &  5.88 & 14376$\pm$49 &      9  \\
  5493 & 18 & 2 & J160337.57+172527.3 & 19.53 & 18.25 & 18.06 & 20.07 & 68403$\pm$63 &  5.47 & \nodata      & \nodata \\
  5522 & 126 & 1 & J160339.26+171105.4 & 16.85 & 15.97 & 15.77 & 17.42 & 10194$\pm$50 &  8.97 & \nodata      & \nodata \\
  5446 & 110 & 1 & J160340.50+172016.6 & 16.50 & 15.93 & 15.56 & 16.59 & 12748$\pm$44 &  6.34 & \nodata      & \nodata \\
\nodata& 47 & 2 & \nodata		&\nodata&\nodata&\nodata&\nodata& 12784$\pm$56 &  5.05 & \nodata      & \nodata \\
  5543 & 39 & 1 & J160340.81+170409.8 & 18.14 & 17.00 & 16.64 & 18.71 & 41139$\pm$47 &  9.36 & \nodata      & \nodata \\
  5688 & 56 & 1 & J160342.84+171812.9 & 18.51 & 17.49 & 17.10 & 18.96 & 29005$\pm$18 & 15.66 & \nodata      & \nodata \\
  5770 & 35 & 2 & J160343.72+171450.8 & 19.50 & 18.59 & 18.07 & 19.80 & 42185$\pm$54 &  6.66 & \nodata      & \nodata \\
  5801 & 106 & 1 & J160344.27+171750.8 & 18.37 & 17.60 & 17.55 & 19.10 & 10694$\pm$42 &  5.93 & \nodata      & \nodata \\
  5827 & 106 & 2 & J160344.58+171719.5 & 18.06 & 17.64 & 17.49 & 18.45 & \nodata      &\nodata& 29142$\pm$18 &      6  \\
  5833 & 95 & 1 & J160344.60+172826.4 & 18.27 & 17.75 & 17.44 & 18.64 & \nodata      &\nodata& 11099$\pm$12 &      7  \\
  5785 & 17 & 1 & J160344.88+170808.1 & 18.21 & 17.60 & 16.53 & 18.10 & \nodata      &\nodata& 20677$\pm$50 &      5  \\
  5855 & 75 & 1 & J160345.04+171347.0 & 17.94 & 17.09 & 16.83 & 18.45 &  9923$\pm$26 & 16.41 & \nodata      & \nodata \\
  5651 & 120 & 1 & J160348.24+171426.2 & 15.31 & 14.47 & 14.04 & 15.63 & 10974$\pm$31 & 12.08 & \nodata      & \nodata \\
  6110 & 83 & 1 & J160350.67+171529.8 & 19.46 & 17.97 & 16.96 & 19.74 & 69091$\pm$48 &  9.80 & \nodata      & \nodata \\
  6065 & 126 & 2 & J160350.71+171421.8 & 17.65 & 16.90 & 15.97 & 17.64 & 12767$\pm$41 &  9.87 & 12855$\pm$58 &     10  \\
  6296 & 73 & 1 & J160354.42+170308.2 & 17.49 & 16.55 & 16.39 & 18.08 & 33321$\pm$33 &  9.11 & \nodata      & \nodata \\
  6331 & 93 & 1 & J160354.55+170435.2 & 18.11 & 17.36 & 17.44 & 18.91 & 10430$\pm$43 &  6.35 & \nodata      & \nodata \\
  6418 & 128 & 1 & J160356.30+170639.3 & 18.46 & 17.70 & 17.48 & 18.96 & 10366$\pm$43 &  8.18 & \nodata      & \nodata \\
  6313 & 122 & 1 & J160356.65+171818.4 & 16.16 & 15.33 & 15.10 & 16.68 & 10004$\pm$45 &  9.80 & \nodata      & \nodata \\
  6472 & 138 & 2 & J160357.52+171419.5 & 18.80 & 17.72 & 17.53 & 19.33 & 41861$\pm$54 &  6.97 & \nodata      & \nodata \\
  6563 & 78 & 1 & J160359.24+173209.4 & 18.46 & 17.43 & 17.13 & 18.96 & 33584$\pm$32 & 13.42 & \nodata      & \nodata \\
  6589 & 102 & 1 & J160359.79+172952.2 & 18.25 & 17.18 & 17.00 & 18.97 & 33642$\pm$45 & 12.30 & \nodata      & \nodata \\
  6569 & 47 & 1 & J160400.07+173104.0 & 17.71 & 16.64 & 16.33 & 18.30 & 33693$\pm$20 & 18.02 & \nodata      & \nodata \\
  6580 & 95 & 2 & J160400.17+173115.6 & 18.85 & 18.00 & 17.38 & 19.27 & 33534$\pm$47 &  9.06 & \nodata      & \nodata \\
  6501 & 20 & 2 & J160400.31+173317.5 & 17.25 & 16.24 & 16.15 & 17.85 & 33757$\pm$41 & 10.55 & \nodata      & \nodata \\
  6678 & 78 & 2 & J160401.12+172839.4 & 18.67 & 17.92 & 17.72 & 19.18 & 34908$\pm$81 &  4.57 & \nodata      & \nodata \\
  6692 & 28 & 1 & J160401.42+170820.4 & 18.55 & 17.91 & 17.68 & 18.95 & 29743$\pm$50 &  4.43 & 29776$\pm$46 &      5  \\
  6512 & 65 & 1 & J160402.75+171656.6 & 15.21 & 14.33 & 14.14 & 15.77 &  9956$\pm$41 & 13.41 & \nodata      & \nodata \\
  7347 & 36 & 1 & J160405.47+171314.7 & 19.36 & 17.70 & 17.61 & 20.02 & 88536$\pm$73 &  5.37 & \nodata      & \nodata \\
  7055 & 131 & 2 & J160407.17+171410.6 & 17.39 & 16.84 & 16.69 & 17.72 & 83790$\pm$118&  2.02 & \nodata      & \nodata \\
  7244 & 96 & 1 & J160407.55+171224.8 & 18.29 & 17.90 & 17.59 & 18.59 & \nodata      &\nodata& 10546$\pm$41 &      7  \\
  7100 & 58 & 1 & J160408.21+173051.1 & 18.58 & 17.75 & 17.61 & 19.11 & 11006$\pm$37 &  9.83 & \nodata      & \nodata \\
  8528 & 40 & 1 & J160418.17+171627.8 & 18.90 & 18.17 &\nodata&\nodata& 10368$\pm$65 &  4.06 & \nodata      & \nodata \\
  8413 & 51 & 1 & J160420.36+172611.2 & 16.67 & 15.90 & 15.65 & 17.07 & 10628$\pm$41 &  8.49 & 10787$\pm$51 &     10  \\
\nodata& 110 & 2 & \nodata		&\nodata&\nodata&\nodata&\nodata& 10805$\pm$68 &  5.51 & 10753$\pm$50 &     10  \\
  8144 & 134 & 1 & J160425.95+172543.1 & 18.86 & 17.85 & 17.60 & 19.43 & 32435$\pm$26 & 13.33 & \nodata      & \nodata \\
  8143 & 17 & 2 & J160427.24+170203.4 & 18.68 & 17.68 & 17.73 & 19.45 & 41147$\pm$83 &  4.41 & \nodata      & \nodata \\
  7584 & 71 & 1 & J160427.99+172148.0 & 17.89 & 16.95 & 16.84 & 18.54 & 32229$\pm$28 & 10.89 & \nodata      & \nodata \\
  8264 & 2 & 2 & J160428.03+171317.1 & 19.20 & 18.38 & 17.74 & 19.08 & 41823$\pm$64 &  2.86 & \nodata      & \nodata \\
  8134 & 2 & 1 & J160428.44+171319.8 & 17.96 & 17.50 & 17.23 & 18.31 & \nodata      &\nodata& 10301$\pm$39 &      6  \\
  7994 & 13 & 1 & J160429.13+172116.8 & 19.11 & 18.32 & 17.58 & 18.96 & \nodata      &\nodata& 53349$\pm$24 &      6  \\
  7895 & 59 & 2 & J160430.42+172809.0 & 15.52 & 14.92 & 14.95 & 16.04 & 11923$\pm$53 &  7.39 & 11831$\pm$27 &     10  \\
  7600 & 51 & 2 & J160434.80+172638.0 & 18.97 & 18.21 & 18.02 & 19.46 &  9692$\pm$63 &  6.94 & \nodata      & \nodata \\
  6780 & 133 & 1 & J160436.01+171739.9 & 18.79 & 17.80 &\nodata&\nodata& 32297$\pm$34 & 10.73 & \nodata      & \nodata \\
  7506 & 44 & 1 & J160437.44+171825.0 & 17.51 & 16.59 &\nodata&\nodata& 32196$\pm$17 & 19.33 & \nodata      & \nodata \\
  5682 & 24 & 1 & J160354.19+173242.7\tablenotemark{a} &\nodata&\nodata&\nodata&\nodata& 15506$\pm$61 &  5.08 & \nodata      & \nodata \\
  3474 & 63 & 2 & J160304.42+171127.6\tablenotemark{a} &\nodata&\nodata& 16.88 & 18.10 & 10871$\pm$75 &  3.00 & 10956$\pm$45 &      9  \\

\enddata
\tablenotetext{a}{No magnitude information is available for this galaxy in the SDSS data.}
\end{deluxetable}
\clearpage

\begin{figure*}[ht!]
\centerline{
\includegraphics[bb=1cm 6cm 20cm 24.3cm,clip=true,width=72mm]{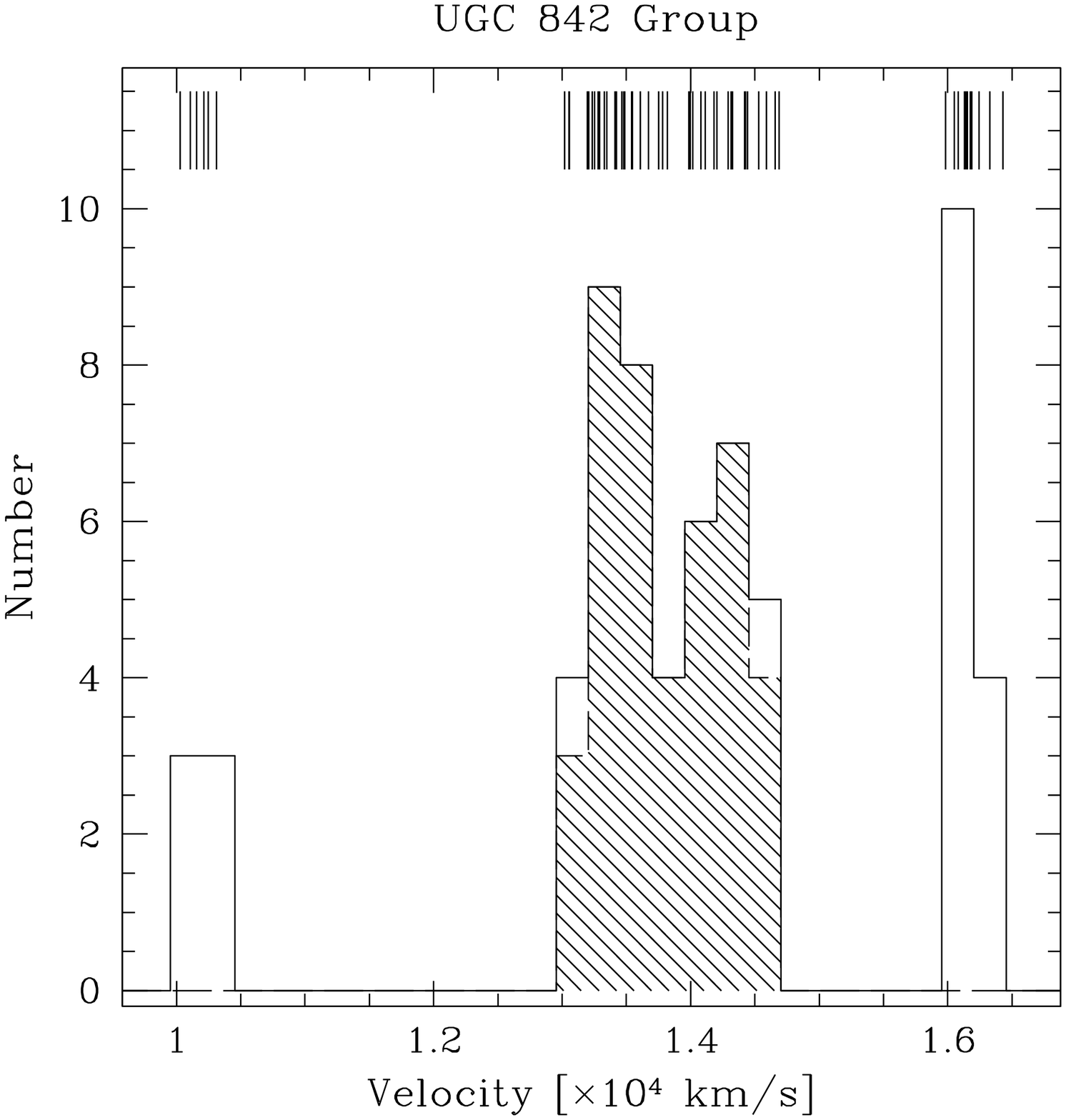}~~~
\includegraphics[bb=1cm 6cm 20cm 24.3cm,clip=true,width=72mm]{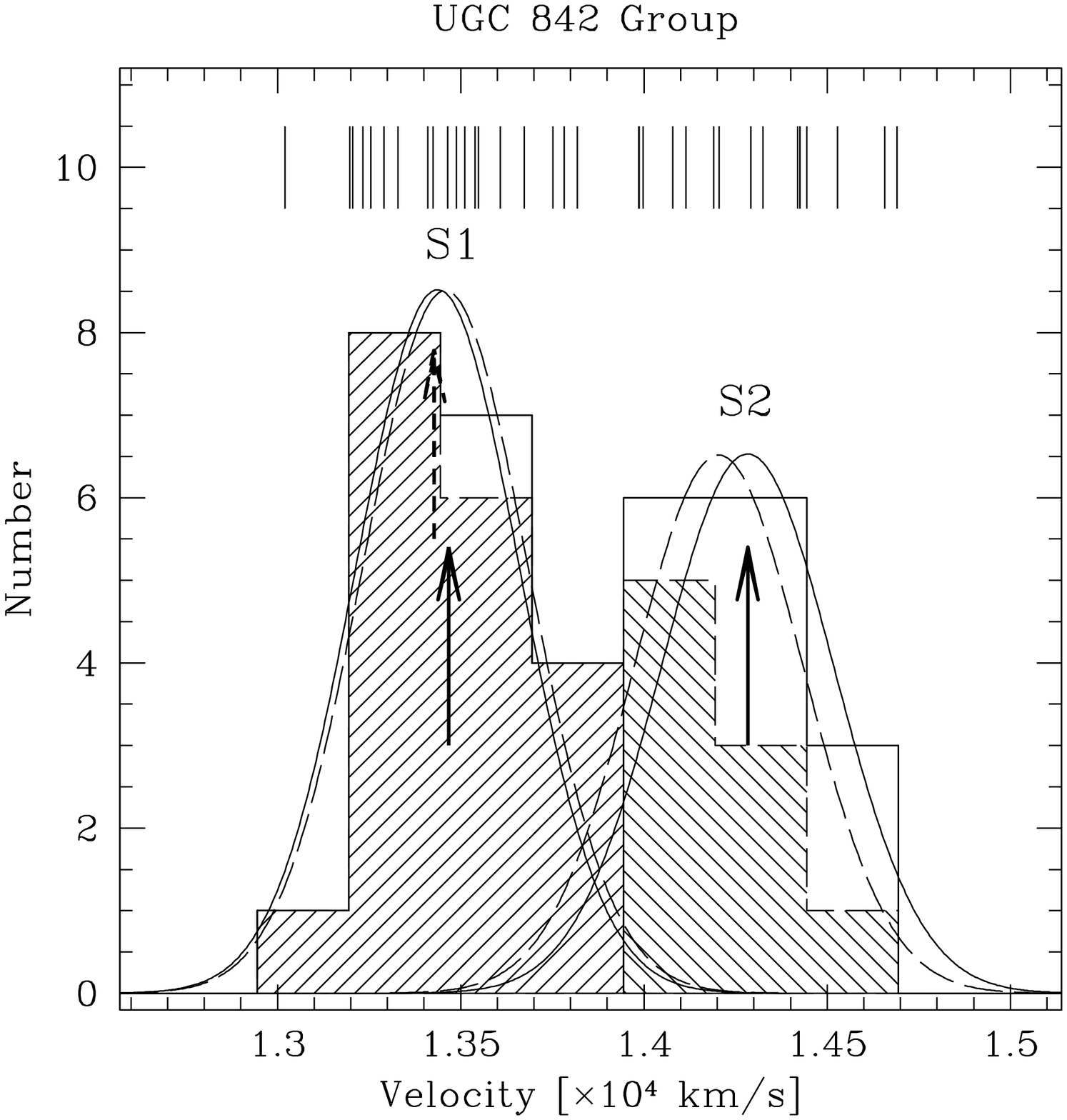}
}
 \caption{
Left: histogram of the velocity distribution of 63 galaxies in the field of UGC 842 with velocities between 10,000 km\,s$^{-1}$ 
and 17,000 km\,s$^{-1}$, inside a region of $\sim$ 80$\times$50 arcmin$^{2}$. The shaded region
indicates the velocity distribution of the potential member galaxies of the group (41 galaxies).
Right: histogram of the velocity distribution for the 35 member galaxies of UGC 842 inside a 20
arcmin ($\sim$ 1 $h_{70}^{-1}$ Mpc) radius. The shaded histogram represents the distribution of non-emission-line galaxies.
The sticks on the upper parts of the boxes represent velocities of individual objects. The arrows indicate the position of the galaxy UGC 842 (dashed line) and the average velocity of the two structures (solid lines).
}
 \label{fig:veldist_ugc}
\end{figure*}

\begin{figure*}[t!]
\centerline{
\includegraphics[bb=0cm 0cm 19cm 20cm,clip=true,width=75mm]{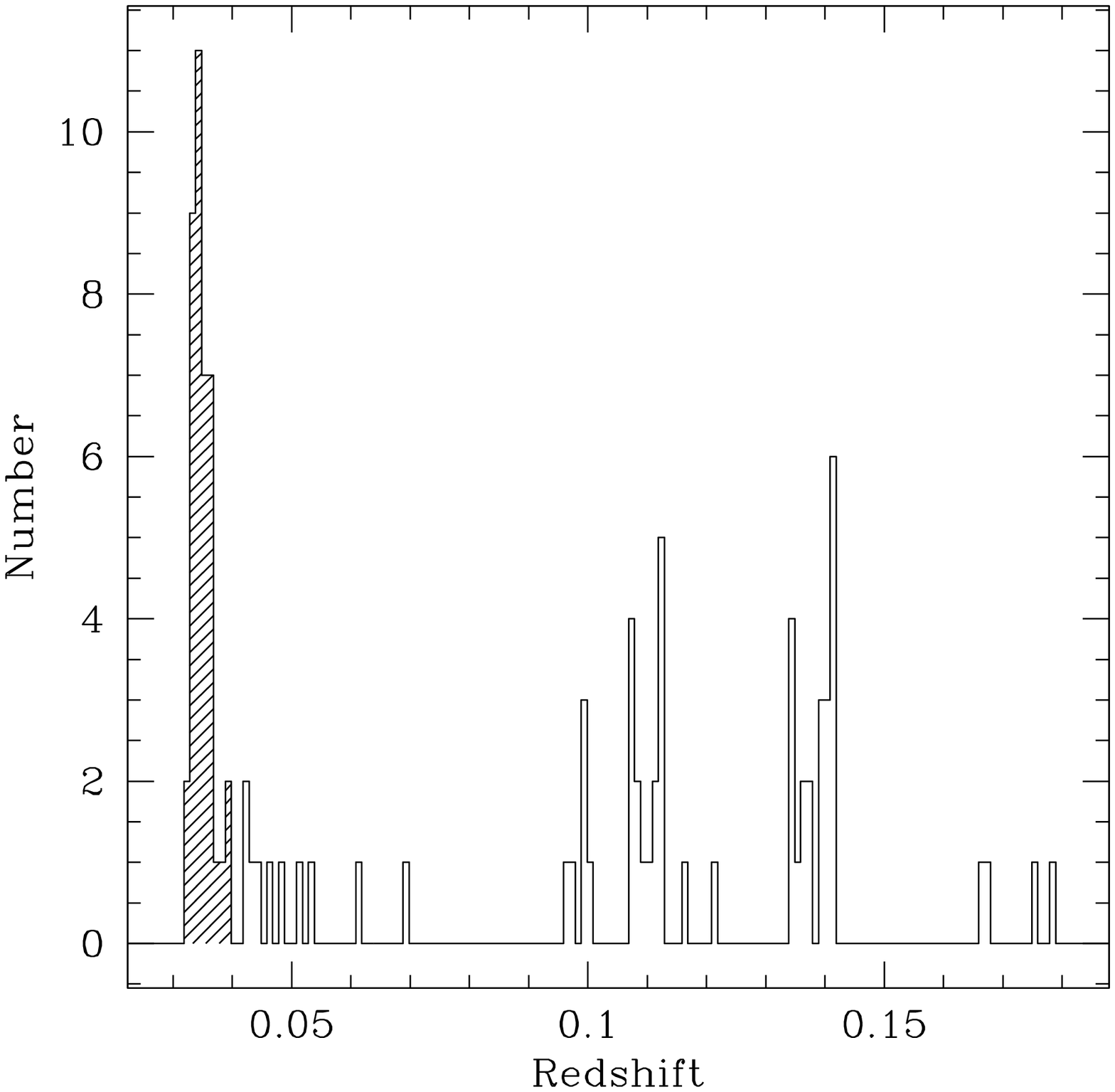}~~~
\includegraphics[bb=0cm 5cm 19.3cm 24cm,clip=true,width=76mm]{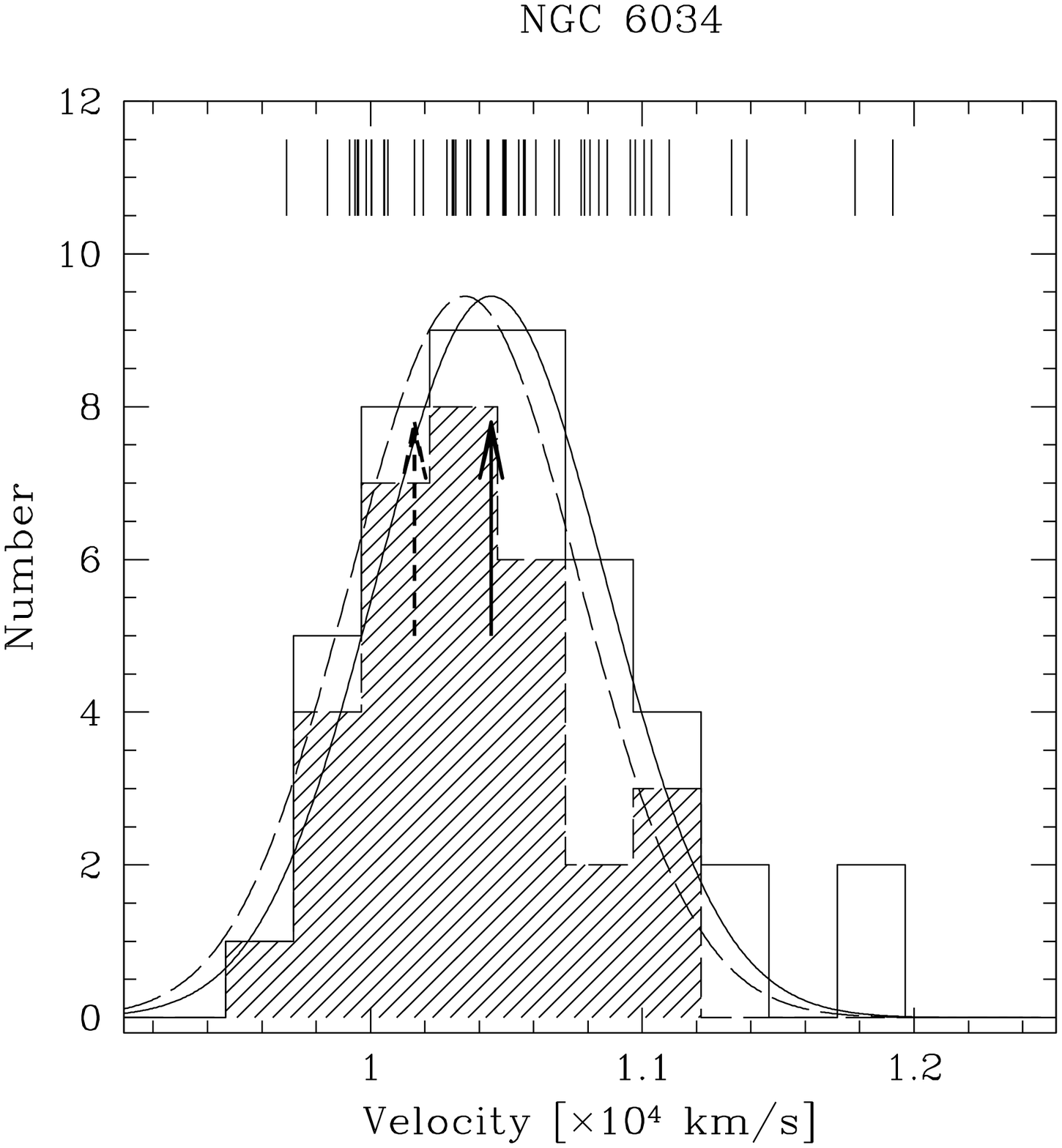}
}
 \caption{
Left: velocity distributions of galaxies with $z$ $<$ 0.2 in the field of NGC 6034. The shaded
histogram, centered at $z$ $\sim$ 0.035, is associated with galaxies at NGC 6034 in the velocity space.
Other background structures are also shown.
Right: histogram of the velocity distribution of member galaxies of NGC 6034 (44, open
histogram) and galaxies members of the group without emission lines (shaded histogram),
and their respective Gaussian profiles. The sticks on the upper parts of the boxes represent velocities of individual objects. The arrows indicate the position of the galaxy NGC 6034 (dashed line) and the average velocity of the distribution (solid line).
}
 \label{fig:veldist}
\end{figure*}

\begin{deluxetable}{lccccccc}
\tabletypesize{\scriptsize}
\tablecaption{Dynamical parameters\label{tab:vel}}
\tablenum{5}
\tablecolumns{8}
\tablewidth{0pc}
\tablehead{
\colhead{Group} &
\colhead{$N_{\rm mem}$} &
\colhead{$<$V$>$ (C$_{\rm BI}$)} &
\colhead{$\sigma_v$ (S$_{\rm BI}$)} &
\colhead{$R_{\rm vir}$} &
\colhead{$M_{\rm vir}$}\\
\colhead{} &
\colhead{} &
\colhead{(km\,s$^{-1}$)} &
\colhead{(km\,s$^{-1}$)} &
\colhead{($h_{70}^{-1}$\,kpc)} &
\colhead{($10^{13}$ $h_{70}^{-1}$\,M$_\odot$)}}
\startdata
UGC 842 (all)            & 35 & 13,808$\pm$81 & 471$\pm$37 & ...                     & ...  \\
UGC 842 (early-type)     & 28 & 13,682$\pm$80 & 419$\pm$51 & ...                     & ...  \\
UGC 842-S1 (all)         & 20 & 13,467$\pm$51 & 223$\pm$29 & 509                     & 1.1  \\
UGC 842-S1 (early-type)  & 19 & 13,458$\pm$53 & 226$\pm$30 & 515                     & 1.1  \\
UGC 842-S2 (all)         & 15 & 14,283$\pm$63 & 235$\pm$35 & ...                     & ...  \\
UGC 842-S2 (early-type)  & 9  & 14,204$\pm$53 & 231$\pm$67 & ...                     & ...  \\

NGC 6034 (all)           & 44 & 10,444$\pm$62 & 410$\pm$39 & 954                     & 7.0    \\
NGC 6034 (non-emission)  & 31 & 10,347$\pm$66 & 361$\pm$40 & 840                     & 4.7  \\

\enddata
\end{deluxetable}

\subsubsection{UGC 842}

The velocity distribution of all known galaxies with velocities between 10,000 km\,s$^{-1}$ and 17,000 km\,s$^{-1}$
and within a region of $\sim$80$\times$50 arcmin${^2}$ is shown in Figure \ref{fig:veldist_ugc} (left). The right histogram in
Figure \ref{fig:veldist_ugc} shows the velocity distribution of the member galaxies of UGC 842 (with projected
distances from the central galaxy within $d_{proj}$ $<$ 1 h$^{-1}_{70}$\,Mpc, or $\sim$ 20 arcmin).
It suggests that UGC 842 has at least a bi modal distribution. In order to investigate its structure, we use the KMM-test \citep{Ashman94}, which is appropriate to detect the presence of two or more components in an observational data set. First we consider whether the data are consistent with a single component. The results of applying the test in the homoscedastic mode (common covariance) yields strong evidence that the velocity distribution of galaxies in the velocity interval above is at least bimodal, rejecting a single Gaussian model at a confidence level of 99.5\% ($P$-value of 0.005). The $P$-value is another way to express the statistical significance of the test, and is the probability that a likelihood test statistic would be at least as large as the observed value if the null hypothesis (one component in this case) were true. 
Assuming two components, the KMM-test estimated a mean value for each component of 13,477 km\,s$^{-1}$ and 14,331 km\,s$^{-1}$. The two components correspond to the structures S1 and S2 in 
Figure \ref{fig:veldist_ugc} (right).  
The average velocities and the velocity dispersions of S1 and S2 calculated using the robust 
bi-weight estimator are listed in Table \ref{tab:vel}. 
The two structures are separated by $\sim$ 820 km s$^{-1}$ in the group rest frame. The first structure includes UGC 842 itself, which has a radial velocity of $\sim$ 13,428 km\,s$^{-1}$, close to the peak of its distribution.
In total, 20 and 15 galaxies are members of the  structures S1 and S2, respectively.
It is interesting to note that the first structure is mainly formed
by passive galaxies (19 out of 20 galaxies)
while for the second one, the number of passive
galaxies is $\sim$ 60\% (9 galaxies) of the total
population.
Similar double peaked galaxy distribution is also found when only 
passive galaxies are considered (the shaded region in Figure \ref{fig:veldist_ugc}, right), which shows that the result is not biased by emission-line objects.
The groups S1 and S2
seem to occupy the same projected area although S1 seems more
centrally concentrated. Although the interaction between two small
groups each with $\sigma_v$ $\sim$ 230 km\,s$^{-1}$ cannot be conclusively discarded, 
a radial velocity difference of $\sim$ 820
km\,s$^{-1}$ between the peaks of the distributions S1 and S2 is an evidence
that the observed bimodality is most probably due to a superposition of structures
in the line of sight. If this is true, and assuming a condition of
equilibrium, the group S1 -- the UGC 842 group -- has a virial mass
and radius of 1.1$\times$10$^{13}$ $h^{-1}_{70}$\,M$_{\odot}$ and 
509 $h^{-1}_{70}$\,kpc, respectively. The high concentration of late-type galaxies in the
group S2 strongly suggests that it is not in equilibrium.

\subsubsection{NGC 6034}

Figure \ref{fig:veldist} (right) shows the velocity distribution of all
known member galaxies (with measured velocities) of the NGC 6034 group.
The distribution is well represented by a Gaussian,
although there is a high velocity tail. 
The best estimates for its parameters are V$_{avg}$ = 10,444$\pm$62 km\,s$^{-1}$ and $\sigma_v$ = 410$\pm$39 km\,s$^{-1}$, from 
44 member galaxies (Table \ref{tab:vel}). This velocity dispersion implies a
virial radius of $\sim$ 954 $h^{-1}_{70}$\,kpc and mass of $\sim$ 7$\times$10$^{13}$ $h^{-1}_{70}$\,M$_{\sun}$ for this group.
The shaded histogram in Figure \ref{fig:veldist} (right) shows the distribution of the
31 non-emission-line (early-type) galaxies of the sample. 
The fraction of emission-line galaxies
in the group is relatively high and represents about $\sim$ 30\% of
the known members, and they  
have preferentially higher velocities than the system,
populating the right tail of the distribution.  For this reason,
we suspect that the tail is due to spiral galaxies falling onto the
principal group. Consequently, we derive a slightly lower dispersion
from the non-emission galaxies ($\sim$ 361 km\,s$^{-1}$) than that of the whole
sample ($\sim$ 410 km\,s$^{-1}$).
If we do not consider the emission-line galaxies, the
calculated average velocity, velocity dispersion, virial radius
and virial mass for the non-emission-line population of NGC 6034 are: $V_{avg}$ =
10,347$\pm$66 km\,s$^{-1}$, $\sigma_v$ = 361$\pm$40 km\,s$^{-1}$,
$R_{virial}$ = 840 $h^{-1}_{70}$\,kpc, and $M_{virial}$ =
4.7$\times$10$^{13}$ $h^{-1}_{70}$\,M$_{\odot}$.  

\begin{figure*}[t!]
\centerline{
\includegraphics[width=80mm]{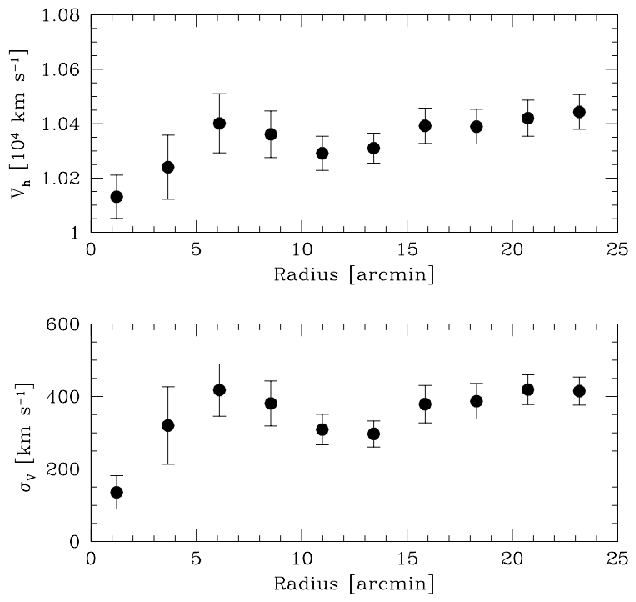}
\includegraphics[width=80mm]{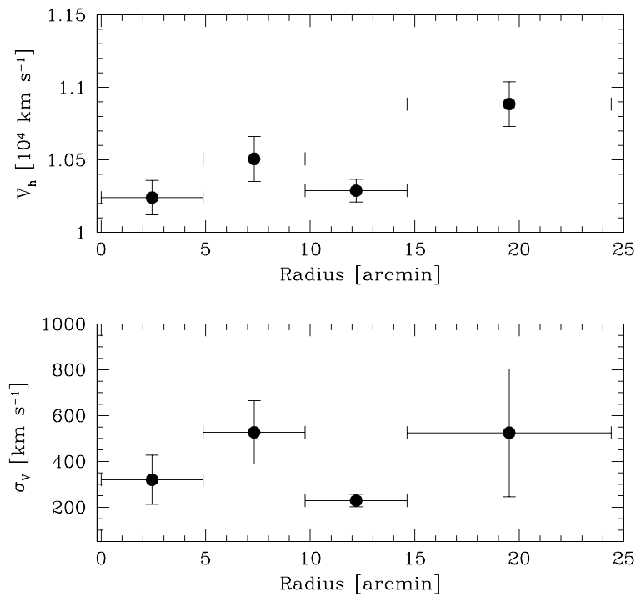}
}
 \caption{
Left: cumulative average velocity and velocity dispersion of galaxies in the field of NGC 6034 as a function of the radius (bins of 100 kpc or 2.44 arcmin).
Right: differential average velocity and velocity dispersion as a function of the radius (rings of 200 kpc or 4.88 arcmin).
}
 \label{fig:cumlvel}
\end{figure*}

In order to check the extent of the NGC 6034 group and to minimize the
contamination of galaxies that could be part of other structures,
we analyzed the cumulative and differential distributions of velocities 
in the group as a function of radius (Figure \ref{fig:cumlvel}).
For the cumulative distribution, we calculated
the average velocity and the velocity dispersion by including all
galaxies inside a given radius, in steps of 100 kpc (denoted in arcmin 
in the figures). 
Thus, for a radius of 400 kpc, we
included all galaxies from the center of the group up to this radius.
With this analysis we can see the general behavior of the group
from the central part to the outskirt regions. For the differential
distribution, we calculated the average
velocity and the velocity dispersion for those galaxies inside 
concentric rings of 200 kpc from the center of the group. If the
ring had less than 10 galaxies, then we increased the width of the
ring until we reached this minimum value. This analysis may reveal 
the degree of anisotropy present in the group.

In both figures we can see an increase in the average velocity and
velocity dispersion for radii larger than 16 arcmin ($\sim$ 650 $h_{70}^{-1}$\,kpc).
This increment probably is due to the presence of the three major
galaxy clusters in the region. We calculated the average
velocity and the velocity dispersion for all galaxies inside this radius. 
The results are $V_{avg}$ = 10,392$\pm$65 km\,s$^{-1}$ and
$\sigma_v$ = 379$\pm$52 km\,s$^{-1}$, which results in $R_{virial}$ $\sim$ 882
$h_{70}^{-1}$\,kpc and $M_{virial}$ $\sim$ 5.5$\times$10$^{13}$ $h_{70}^{-1}$\,M$_{\odot}$.
These are the values we would favor for describing the group NGC 6034. Even with this 
revised smaller $R_{virial}$ for NGC 6034, three galaxies 
(J160348.24+171426.2, J160402.75+171656.6, and J160356.65+171818.4)
populate the $\Delta r$ $<$ 2 magnitude gap, within $R_{virial}$/2 of the group center.

\subsubsection{Optical Versus X-ray Properties}
\label{sct:cons}

The measured value of the intracluster gas bolometric X-ray luminosity of NGC 6034 is $\sim$ 2.8$\times$10$^{43}$ $h_{50}^{-2}$
erg\,s$^{-1}$ \citep{Fukazawa04} and for UGC 842 it is $\sim$ 1.63$\times$10$^{43}$\,$h^{-2}_{71}$\,erg\,s$^{-1}$ \citep{Voevodkin08}. 
According to the relations for groups and clusters from \citet{Xue00} and \citet{Mahdavi01}, 
we find that for the X-ray luminosity of NGC 6034 group it is expected
a velocity dispersion of about 425 km\,s$^{-1}$, in good agreement 
with our direct measurement of 410 km\,s$^{-1}$ (when we include all galaxies
in the group; or 360 km\,s$^{-1}$ when only the non-emission-line galaxies are
included).
On the other hand, a velocity dispersion of $\sim$ 450 km\,s$^{-1}$ is expected for a group with the X-ray luminosity of UGC 842, 
which is much higher than its measured
velocity of 223 km\,s$^{-1}$ even  
if we assume that each substructure seen in its line of sight 
contributes with one-half of the observed L$_{X}$.
The first consequence of this divergence between X-rays and optical properties
is the determination of the true virial radius for UGC 842. 
From dynamical relations of velocity distribution
of galaxies we determine a virial radius of 509 $h^{-1}_{70}$\,kpc, while the intragroup X-ray gas 
implies a virial radius of 1272 $h^{-1}_{70}$\,kpc, according to \citet{Gastaldello07}.
This apparent inconsistency is discussed in Section \ref{sct:disc}.

\section{Summary and Discussion}
\label{sct:disc}

We summarize below the main findings of this paper:

About NGC 6034:

\begin{itemize}

\item
NGC 6034 is a group of $\sim$ 4 L$^{*}$ galaxies with a mass 
of $\sim$ 7$\times$10$^{13}$ $h^{-1}_{70}$\,M$_{\odot}$ and $R_{virial}$ of $\sim$ 954 $h^{-1}_{70}$\,kpc 
(or $\sim$ 4.7$\times$10$^{13}$ $h^{-1}_{70}$\,M$_{\odot}$ and $R_{virial}$ of $\sim$ 840 $h^{-1}_{70}$\,kpc
if only non-emission-line galaxies are considered). 
It is not a fossil group given that the magnitude difference in the $r$ band between the first
and second (J160402.75+171656.6) ranked galaxies is only 0.79 mag and the projected distance between these two galaxies is $\sim$ 360 $h_{70}^{-1}$\,kpc. Other four (or five; see Section \ref{sct:cmd}) galaxies also violate the optical criteria to classify it as a fossil group. 

\item 
NGC 6034 is clearly part of a much larger
structure that includes at least three clusters (among them 
the Hercules cluster) and several other groups.

\item The velocity distribution of NGC 6034 is fairly well represented by
a Gaussian except for a high velocity tail composed of spiral galaxies.
These are most probably objects falling onto the system. 
The velocity dispersion of
only the non-emission-line galaxies (31 members with measured redshifts)
is 361$\pm$40 km\,s$^{-1}$ and including the 13 emission-line objects it increases 
slightly to 410$\pm$39 km\,s$^{-1}$.

\end{itemize}

About UGC 842:

\begin{itemize}

\item The system referred to as UGC 842 in the literature is in fact 
a superposition of two groups, S1 and S2, with a velocity difference of 
about 820 km\,s$^{-1}$.

\item UGC 842/S1 is dominated by passive galaxies
while there is a high fraction (40\%) of emission galaxies in S2.

\item UGC 842/S1 is dominated by a bright elliptical galaxy and it is a low mass fossil {\it group}.
The large content in passive galaxies suggests equilibrium, and the estimated virial mass and radius
are 1.1$\times$10$^{13}$ $h^{-1}_{70}$\,M$_{\odot}$ and 509 $h^{-1}_{70}$\,kpc, respectively. 

\item There is a large discrepancy between the expected temperature and
X-ray luminosity expected for a group with such a low sigma and its  
observed values of $kT$ of 1.90$\pm$0.30\,keV and
L$_{X,\rm bol}$ $\sim$ 1.63$\times$10$^{43}$\,$h^{-2}_{71}$ erg\,s$^{-1}$ by \citet{Voevodkin08}, suggesting that we have
a case of interaction of the two subclumps S1 and S2 
or we are whitenessing the
decrease of sigma due to the central merger event.

\end{itemize}

Other previous papers have also studied UGC 842. However, it has not been realized in these works that what is referred to as UGC 842 is in fact two systems. For example, a velocity dispersion of 439 km\,s$^{-1}$ was derived by \citet{Voevodkin08} from the redshifts of 16 galaxies within a radius of 509 $h^{-1}_{71}$\,kpc around UGC 842. This value is in good agreement with our value of 471$\pm$37 km\,s$^{-1}$ determined from the redshifts of 35 potential member galaxies, if we ignore the double-peaked velocity distribution. 
However, the KMM-test \citep{Ashman94} rejects a Gaussian velocity distribution at the 99.5\% level, and the distribution reveals clearly two structures -- S1, the UGC 842 group, and S2 -- with a velocity dispersion of about 230 km\,s$^{-1}$ each (see Figure \ref{fig:veldist_ugc} and Table \ref{tab:vel}). 
The fact that the bolometric X-ray luminosity of UGC 842 fits well in the L$_X$-T$_X$ relation, the measured sigma for the group is too low in the L$_X$--$\sigma_v$ and T$_X$--$\sigma_v$ group relations \citep[e.g.,][]{Xue00}: 
UGC 842 seems to be too luminous and too hot, with a $kT$ of $\sim$\,1.9 keV \citep{Voevodkin08}, for its sigma.
The expected X-ray temperature for S1 and S2 from their velocity 
dispersion is around 0.5--1 keV, and the superposition of both plasmas in the line of sight does not imply in the detection of a temperature as high as $kT$\,$\sim$\,1.9\,keV.
Although not too pronounced, similar behavior has also been observed in other fossil groups \citep{Khosroshahi07}.
In fact, this high value for the temperature of UGC 842 is in line with the 
suggestion of \citet{Khosroshahi07} that both temperature and X ray luminosity has been boosted for fossil groups compared to the values for ``normal'' groups with similar velocity dispersions, given their early times of formation. 
Another possibility we envisage is that the galaxies have lower relative 
velocities in fossil groups given that they have lost energy by dynamical 
friction in the process of spiraling towards the group center for interacting 
and finally merging.
The radial velocity difference between the peaks of the velocity distribution of S1 and S2 
($\sim$ 820 km\,s$^{-1}$; see Section \ref{sct:veldist}) suggests that the 
two groups are simply overlapping in the line of sight. Although the smooth appearance of the X-ray image \citep{Voevodkin08} agrees with this hypothesis, we think the SNR of the X-ray observation cannot rule out the presence of X-ray
substructures that could indicate a recent interaction of the two sub-groups.
Better S/N and spatial resolution X-ray data may reveal if there is an interaction
between the two groups (S1 or S2) or not. 
Thus, the measured temperature of $kT$ $\sim$ 1.9 keV for UGC 842 
may either represent the intragroup medium of the most massive group, S1, or it could be the 
result of an interaction between S1 and S2. 	
Both possibilities make UGC 842 especially interesting in the study of formation and evolution of fossil groups.

\acknowledgments

We thank financial support from the Brazilian agency FAPESP (Funda\c c\~ao de Amparo \`a Pesquisa do Estado de S\~ao Paulo) and CNPq (Conselho Nacional de Desenvolvimento Cient\'ifico e Tecnol\'ogico). R.L.O.: FAPESP Postdoctoral Research Fellow grant -- number 2007/04710-1.
E.R.C. is supported by the Gemini Observatory, which is operated by the Association of Universities
for Research in Astronomy, Inc., on behalf of the international Gemini partnership of Argentina,
Australia, Brazil, Canada, Chile, the United Kingdom, and the United States of America.
We thank Renato Dupke for valuable discussions on X-ray properties of groups/clusters of galaxies.
This research has made use of the NASA/IPAC Extragalactic Database (NED) which is operated by the Jet Propulsion Laboratory, California Institute of Technology, under contract with the National Aeronautics and Space Administration, and use of data of the Sloan Digital Sky Survey.
Funding for the SDSS and SDSS-II was provided by the Alfred P. Sloan Foundation, the Participating Institutions, the National Science Foundation, the U.S. Department of Energy, the National Aeronautics and Space Administration, the Japanese Monbukagakusho, the Max Planck Society, and the Higher Education Funding Council for England. The SDSS was managed by the Astrophysical Research Consortium for the  Participating Institutions.

\end{document}